\documentclass[floatfix,aps,amsmath,prd,twocolumn,notitlepage,eqsecnum,showpacs,nofootinbib,superscriptaddress]{revtex4-1}
\usepackage{hyperref,url}
\usepackage{graphicx,graphics,amsfonts,amsmath,mathrsfs,amssymb,bm,psfrag,natbib,dcolumn,textcase}
\usepackage{array}
\usepackage{enumerate}
\usepackage{grffile}
\usepackage[utf8]{inputenc}
\usepackage[normalem]{ulem}
\usepackage{verbatim}
\allowdisplaybreaks
\usepackage{amsmath}

\usepackage[english]{babel}
\usepackage[usenames]{color}
\definecolor{darkgreen}{rgb}{0,0.5,0}
\usepackage{subfigure}
\usepackage{bigints}
\usepackage{comment}
\usepackage{float}
\usepackage{multirow}
\usepackage{placeins}
\usepackage{hyperref}
\usepackage{xspace,ulem}
\usepackage{graphicx,float,xcolor}
\usepackage{dcolumn}
\usepackage{bm,amsmath, amssymb,}
\usepackage[utf8]{inputenc} 
\usepackage{dblfloatfix} 
\usepackage{graphics}
\usepackage{subfigure}
\usepackage[capitalise]{cleveref}
\usepackage{comment}

\begin{document}
\title{Systematic bias on parametrized tests of general relativity\\ due to neglect of orbital eccentricity} 
\date{\today}
\author{Pankaj Saini}
\email{pankajsaini@cmi.ac.in}
\affiliation{Chennai Mathematical Institute, Siruseri 603103, India}
\author{Marc Favata}
\email{marc.favata@montclair.edu}
\affiliation{Department of Physics \& Astronomy, Montclair State University, 1 Normal Avenue, Montclair, New Jersey 07043, USA}
\author{K.~G.~Arun}
\email{kgarun@cmi.ac.in}
\affiliation{Chennai Mathematical Institute, Siruseri 603103, India}
\begin{abstract}
Gravitational-wave observations provide a unique opportunity to test general relativity (GR) in the strong-field and highly dynamical regime of the theory. Parametrized tests of GR are one well-known approach for quantifying violations of GR. This approach constrains deviations in the coefficients of the post-Newtonian phasing formula, which describes the gravitational-wave phase evolution of a compact binary as it inspirals. Current bounds from this test using LIGO/Virgo observations assume that binaries are circularized by the time they enter the detector frequency band. Here, we investigate the impact of residual binary eccentricity on the parametrized tests. We study the systematic biases in the parameter bounds when a phasing based on the circular orbit assumption is employed for a system that has some small residual eccentricity. We find that a systematic bias (for example, on the leading Newtonian deformation parameter) becomes comparable to the statistical errors for even moderate eccentricities of $\sim 0.04$ at $10$ Hz in LIGO/Virgo band for binary black holes, and $\sim 0.008$ for binary neutron stars. This happens at even lower values of orbital eccentricity in the frequency band of third-generation (3G) detectors like Cosmic Explorer ($\sim 0.005$ at $10$ Hz for binary black holes and $\sim 0.002$ for binary neutron stars). These results demonstrate that incorporating physical effects like eccentricity in waveform models is important for accurately extracting science results from future detectors.
\end{abstract}
\maketitle

\section{Introduction}
General relativity (GR) has been tested in weak-field and strong-field regimes via various experiments \cite{Will:2005va, Sathyaprakash:2009xs,Yunes:2013dva,Berti:2015itd,Krishnendu:2021fga}, with no deviations found to date \cite{GW150914TGR:2016lio,GWTC1TGR:2019fpa,GW170817TGR:2018dkp,GWTC2:2020tif,GWTC3:2021sio,Kramer:2021jcw}. Gravitational waves (GWs) probe the dynamical and strong field regime of GR, where nonlinear effects play a dominant role. Waveform models including the inspiral, merger, and ringdown of compact objects are now well developed in GR. Such models are less developed in alternative gravity theories, particularly for the highly nonlinear merger and ringdown phases \cite{Sennett:2016klh,Bernard:2018hta,Bernard:2018ivi,Tahura:2018zuq,Khalil:2018aaj,Shiralilou:2021mfl,Okounkova:2019zjf,Okounkova:2020rqw,Cayuso:2020lca,East:2020hgw,East:2021bqk}. Considering the historical difficulty of solving the two-body and waveform modeling problems in GR, it may be impractical to also solve those problems for large numbers of viable alternative theories. An alternative approach is to test GR by comparing GW signals to waveform templates that differ from GR predictions via small parametrized deviations \cite{Yunes:2009ke,TIGER:2013upa}.

Additional fields or higher curvature corrections change the dynamics of compact binaries and their radiation from the GR predictions \cite{Yagi:2011xp,Lang:2013fna,Lang:2014osa,Bernard:2018hta,Bernard:2018ivi,Julie:2018lfp}. For example, the presence of dipole radiation in some alternative theories \cite{Will:1994fb,Will:2004xi} alters the GW flux from the binary, leading to a faster decay of the binary's orbit. These effects are imprinted on the observed GW phasing from the binary. One well-developed approach to test GR using GWs is the {\it parametrized tests of GR} (TGR) framework \cite{Arun:2006hn,TIGER:2013upa,Arun:2004hn,PPE:2011ys}. In this approach the inspiral part of the frequency-domain GW phase is parametrized in terms of non-GR deformation parameters. These non-GR deformation parameters are introduced as free parameters at each post-Newtonian (PN) order  \cite{Datta:2020vcj} and attempt to capture  a particular class of deviations from GR. The value of these deformation parameters is zero in GR, hence these tests are \emph{null tests}. Potential deviations would accumulate over many GW cycles in a detected GW signal. Precise measurement of these non-GR deformation parameters provide a constraint on potential GR deviations. (For a recent overview and comparison of multiple approaches for testing GR, see Ref.~\cite{nathan-GRtestsPRD2022}.)

As GWs carry away energy and angular momentum from a binary, an initially elliptical orbit tends to circularize \cite{PetersMathews:1963ux,Peters:1964zz}. In the small eccentricity limit, the eccentricity decreases approximately inversely to the increasing GW frequency, $e_t/e_{0}  \approx (f_{0}/f)^{19/18}$ \cite{Peters:1964zz}. Here $e_t$ is the binary eccentricity when the GWs at twice the orbital frequency have frequency $f$, and $e_0$ is the value of $e_t$ at a reference frequency $f_0$.\footnote{Formally, the subscript $t$ in $e_t$ denotes the ``time-eccentricity'' parameter in the quasi-Keplerian formalism \cite{DGI}. It also emphasizes that quantity's time evolution and distinguishes it from Euler's number.} Eccentricity thus decays rapidly. For example, a binary with moderate initial eccentricity of $0.2$ at $0.1$ Hz reduces to one with eccentricity $\sim$ $10^{-3}$ when it reaches a GW frequency of $10$ Hz (the lower limit of the LIGO frequency band). Present bounds on non-GR deformation parameters \cite{GWTC3:2021sio} assume that binaries are quasicircular when entering the detectors' frequency band (and are well modelled by circular waveform templates). 

{\it The main purpose of this paper is to evaluate the effect on parametrized tests of GR if the binary is in fact eccentric, but modeled using circular waveforms.} The unmodeled binary eccentricity will potentially bias our parameter estimation. Understanding the extent of this bias on the parameters that model GR violations is essential for performing a meaningful test of GR. This will be especially important for future third generation (3G) GW detectors that will operate with higher levels of sensitivity. 

The role of systematic errors in GR tests is also addressed in several other works. For example, Ref.~\cite{Moore:2021eok} considers how systematic bias on GR tests can accumulate in a large catalog of GW events. Reference \cite{Pang:2018hjb} showed how the neglect of higher modes in the waveform model could bias GR tests. Reference~\cite{Narikawa:2016uwr} considered the role of unmodeled orbital eccentricity, spin, and tidal deformation as possible sources of systematic bias using the parametrized post-Einsteinian formalism \cite{Yunes:2009ke}.

The expected eccentricity of inspiralling compact objects is dependent on how these binaries formed. There are mainly two formation channels for compact objects: the isolated formation channel and the dynamical formation channel \cite{Postnov:2014tza,Mapelli:2021taw}. The isolated channel is likely to produce binaries with low to moderate eccentricities. But the dynamical formation of compact objects in dense environments like globular clusters (GCs) and nuclear clusters can form highly eccentric binaries \cite{Samsing:2017xmd}. When these sources enter the frequency range of ground-based detectors, they could still possess non-negligible orbital eccentricity \cite{Antonini:2013tea,OLeary:2008myb,Antonini:2012ad,Wen:2002km}.

Based on N-body simulations of GCs, Ref.~\cite{Antonini:2013tea} showed that $\sim$ $50\%$ of the coalescing black hole (BH) binaries driven by the Kozai mechanism in GCs will have eccentricities $>0.1$ at $10$ Hz. Reference \cite{Antonini:2012ad} indicates that about $0.5\%$ of merging compact binaries in galactic nuclei near supermassive BHs (SMBHs) will enter the LIGO band with eccentricity $\geq$ $0.5$. Reference \cite{OLeary:2008myb} studied the stellar mass BHs that form density cusps due to mass segregation near SMBHs. They predict that most binaries ($\sim$ $90\%$) that form through BH scattering in such dense environments will have eccentricities  $>0.9$ when they enter the LIGO band. Reference \cite{Wen:2002km} studied hierarchical triple BH systems in GCs, predicting that $\sim$ $30\%$ of those systems will have eccentricities $>0.1$ at $10$ Hz. Work in \cite{zevin-etal2021} estimated that $\sim 4\%$ of the binary black holes (BBHs) detectable from clusters will have measurable eccentricity. Recent claims have indicated signatures of (or constraints on) binary eccentricity in observed GW events, suggesting a subpopulation of dynamically formed binaries \cite{Romero-Shaw:2020thy,Gayathri:2020coq,Romero-Shaw:2021ual,OShea:2021ugg}. See Sec.~IA of \cite{Favata:2021vhw} for additional discussion and references on the expected eccentricities of compact objects in the frequency band of ground-based detectors. 

The current sensitivities of GW detectors make it difficult to put strong constraints on orbital eccentricity if the eccentricity is $\lesssim 0.01$ in the detector's frequency band \cite{Favata:2021vhw} (see also \cite{Lower:2018seu}). However, future 3G detectors can more tightly constrain the orbital eccentricity. Precise measurement of eccentricity will provide crucial information about the formation of compact object binaries \cite{Nishizawa:2016jji}.  

The detection of eccentric binaries in ground-based detectors using quasicircular waveforms has been studied in Refs.~\cite{Huerta:2013qb,Sun:2015bva, LIGOScientific:2019dag,Ramos-Buades:2020eju}. They reported that binaries with eccentricities $\lesssim 0.02$--$0.15$ would be detectable using quasicircular waveform templates. Moreover, parameter estimation of eccentric binaries using quasicircular waveforms may lead to significant systematic biases in the estimated parameters \cite{Favata:2013rwa}. Using the formalism developed in \cite{Cutler:2007mi}, Favata \cite{Favata:2013rwa} studied systematic biases in the context of binary neutron stars (BNS) in LIGO and 3G detectors, considering the neglect of high-PN-order terms, spins, eccentricity, and tidal deformation. He found that even small eccentricities ($e_{0} \sim 10^{-3} \mbox{--} 10^{-2}$) can lead to non-negligible systematic biases, with systematic errors exceeding statistical errors in some regions of parameter space. Systematic errors are independent of the signal-to-noise ratio (SNR), while statistical errors scale as ${\rm SNR}^{-1}$. Hence, for the high SNR sources expected in 3G detectors, systematic errors could easily dominate the statistical errors \cite{Berti:2006ew}. 

A recent study \cite{Favata:2021vhw} has shown that systematic biases due to binary eccentricity become significant when binaries enter the LIGO band with an initial orbital eccentricity of $e_{0}$ $\geq$ $0.01$ to $0.1$. These systematic biases will affect the estimation of intrinsic parameters of the binary (masses and spins). Similarly, neglecting orbital eccentricity in the waveform can also lead to systematic biases in tests of GR that could mimic GR violations.  This will be a more serious issue for future 3G detectors such as Cosmic Explorer (CE) \cite{CE:2019iox}  and the Einstein Telescope~\cite{ET:2016wof,Maggiore_ET_Science_Case:2019uih}. In addition to their improved sensitivities at all frequencies, the excellent low-frequency sensitivity of 3G detectors means that binaries may retain more residual eccentricity in the frequency band of 3G detectors relative to current (2G) detectors. In this paper, we quantify the effect of systematic errors due to the neglect of orbital eccentricity on the estimation of the TGR deformation parameters.

To investigate this issue we adapt the analysis in \cite{Favata:2021vhw} to include the TGR parameters. The standard PN waveform in the frequency domain ({\tt TaylorF2}) for circular binaries is adopted, modified by the TGR deformation parameters. We then make use of the {\tt TaylorF2Ecc} model \cite{Moore:2016qxz} for low-eccentricity binaries to act as the ``source'' of the systematic bias. 
Section ~\ref{section2} of this paper describes the waveform model in more detail. To calculate the statistical  errors on the TGR parameters, we apply the Fisher information matrix formalism. Systematic parameter errors are computed via the Cutler and Vallisneri formalism \cite{Cutler:2007mi}. Both formalisms are summarized in Sec.~\ref{section3}. Our results and conclusions are presented in Secs.~\ref{section4} and \ref{sec:conclusion}.

\section{Waveform model}\label{section2}
GWs have two independent polarization states in GR: plus and cross. The GW signal $h(t)$ measured in the detector is a linear combination of the amplitudes $h_{+,\times}$ of these two polarization states, 
\begin{equation}\label{strain}
    h(t) = F_{+} h_{+}(t) + F_{\times} h_{\times}(t) \,.
\end{equation}
Here $F_{+,\times}$ are the antenna pattern functions of the detector, which depend on the sky location of the source and a polarization angle $\psi$.

In the stationary phase approximation (SPA), the Fourier transform of $h(t)$ can be expressed in the form
\begin{equation}\label{waveform}
    \Tilde{h}(f) = \mathcal{A} e^{i \Psi(f)} = \hat{\mathcal{A}}f^{-7/6} e^{i \Psi(f)} \,.
\end{equation}
Averaging over the antenna pattern functions and the inclination angle, the amplitude parameter $\hat{\mathcal A}$ in the quadrupole approximation is given by 
\begin{equation}\label{amplitude}
  \hat{\mathcal{A}} = \frac{1}{\sqrt{30}\pi^{2/3}} \frac{\sqrt{\eta} M^{5/6} (1+z)^{5/6}}{d_{L}} \,.
\end{equation}
Here $\eta=m_{1}m_{2}/M^{2}$ is the symmetric mass ratio, $m_{1,2}$ are the component masses in the source frame, $M=m_1+m_2$ is the total mass, $d_{L}$ is the luminosity distance to the source, and $z$ is the source redshift. We use the cosmological parameters for a flat universe given in Ref.~\cite{Planck:2015fie}:
$H_{0}=67.90$(km/s)/Mpc, $\Omega_{m}=0.3065$, and $\Omega_{\Lambda}=0.6935$.

In PN theory \cite{Blanchet:1995ez,Blanchet:1995fg,Kidder:1995zr,Blanchet:2002av,Blanchet:2006gy,Arun:2008kb,Marsat:2012fn,Mishra:2016whh}, the SPA phase $\Psi(f)$ for circular binaries is expanded in powers of the relative orbital velocity parameter $v = [\pi M  (1+z) f]^{1/3}$: 
\begin{equation}
\label{phase}
     \Psi(f)  =  2\pi ft_{c} + \phi_{c} + \frac{3}{128 \eta v^{5}} \sum_{k} (\varphi_{k}v^{k} 
      +  \varphi_k^{\rm log}  v^{k} \ln v) \,,
\end{equation}
where $t_{c}$ and $\phi_{c}$ are the time and phase of coalescence respectively. Different powers of $v$ relative to the leading-order (Newtonian) term [$\mathcal{O}(1/v^5)$] denote different PN orders. In our terminology, corrections of order $v^{k}$ relative to the Newtonian term correspond to $k/2$ PN-order corrections. The PN coefficients $\varphi_{k}$ with the summation index varying over $k = (-2,0,2,3,4,6,7)$ denote the PN corrections up to 3.5PN order. A constant term (independent of frequency) at $2.5$PN order $(k=5)$ is absorbed into a redefinition of $\phi_{c}$.\footnote{This redefinition, $\phi_c + \frac{3}{128\eta}\varphi_5 \rightarrow \phi_c$, changes the $\eta$ dependence of $\Psi(f)$ and the interpretation of the coalescence phase $\phi_c$. The choice to absorb this constant-frequency 2.5PN phase correction into a redefined $\phi_c$ can have non-negligible impacts on the values of the statistical and systematic parameter errors computed in Sec.~\ref{section3}, via changes in the derivatives $\partial h/\partial \theta^a$ for $\theta^a=\eta$. When performing a statistical best fit, the inferred parameter values of a binary depend on the set of parameters that are allowed to vary as well as the waveform model that is adopted. Redefinitions of $\phi_c$ effectively constitute a change of the waveform.} The negative PN term $(k =-2)$ corresponds to a correction in the GW phasing produced by dipole radiation; this term is zero in GR but nonzero in some alternative theories of gravity \cite{Will:1994fb,Will:2004xi}. The coefficients $\varphi_k^{\rm log}$, with the index varying over $k = (5,6)$, multiply the PN corrections associated with a logarithmic frequency dependence. 

Each PN coefficient is a function of the intrinsic parameters of the compact binary, such as the symmetric mass ratio $\eta$ and the dimensionless spin parameters $\chi_{1,2}$. (In our analysis we assume that the compact-object spins are aligned or antialigned with the orbital angular momentum of the binary.) If the source parameters $\eta$, $\chi_1$, and $\chi_2$ are determined, the values of the PN coefficients $(\varphi_k, \varphi_k^{\rm log})$ are known in GR. To 3.5PN order, the values of $(\varphi_k^{\rm GR}, \varphi_k^{\rm GR,\, log})$ can be read off of equations in Refs.~\cite{Arun:2004hn,Arun:2008kb,Buonanno:2009zt,Wade:2013hoa,Mishra:2016whh}. Spin-orbit, spin-spin and self-spin terms are included to 3.5PN order; quadrupole-monopole terms are also included for BBHs. For binary neutron stars, finite-sized effects are ignored. Deviation of a coefficient from the GR value is a potential signature of a different underlying theory of gravity.

To parametrize deviations in the PN coefficients, we modify the waveform by introducing \emph{TGR parameters} at each PN order:
\begin{subequations}
\label{deformation}
\begin{align}
\varphi_{k} &\xrightarrow{} \varphi_{k}^{\rm GR} (1 + \delta \hat{\varphi}_{k}) \,, \\
\varphi_k^{\rm log} &\xrightarrow{} \varphi_k^{\rm GR,\, log} (1 + \delta \hat{\varphi}_k^{\rm log}) \, .
\end{align}
\end{subequations}
Here $\delta \hat{\varphi}_{k}$ and $\delta \hat{\varphi}_k^{\rm log}$ are the fractional TGR deformation parameters. By definition their values are zero in GR. Since the $-1$PN term vanishes in GR ($\varphi^{\rm GR}_{-2}=0$), the deviations are absolute for this term ($\varphi_{-2}\rightarrow \delta \hat{\varphi}_{-2}$). The precise measurement of these TGR parameters is an important tool for searching for potential GR deviations. 

To estimate the values of the source and TGR parameters under the assumption of a circular binary, we can apply the waveform described by Eqs.~\eqref{phase} and \eqref{deformation}. To model a systematic bias in those parameters due to orbital eccentricity, we modify the waveform SPA phase via
\begin{equation}
\label{decomposed phase}
    \Psi(f) \rightarrow \Psi(f)^{\rm circ., TGR} + \frac{3}{128 \eta v^{5}} \Delta\Psi^{\rm ecc.}_{\rm 3PN} \,.
\end{equation}
Here $\Psi(f)^{\rm circ., TGR}$ refers to the circular PN phasing including spin and TGR parameters [Eqs.~\eqref{phase} and \eqref{deformation}]. The $\Delta\Psi^{\rm ecc.}_{\rm 3PN}$ term arises from the {\tt TaylorF2Ecc} waveform approximant developed in Ref.~\cite{Moore:2016qxz}.
That waveform is an extension of the standard circular PN {\tt TaylorF2} frequency-domain approximant, but includes leading-order [$\sim \mathcal{O}(e_0^{2})$] corrections due to eccentricity in the GW phasing to 3PN order. Eccentric corrections to the circular waveform amplitude are neglected. Since GW detectors are more sensitive to the GW phase than to the amplitude, small eccentric corrections to the amplitude will be less important than corrections to the phase. 

The structure of $\Delta\Psi^{\rm ecc.}_{\rm 3PN}$ can be understood from the following expression:
\begin{align}
    \label{eccentric phase}
    \Delta\Psi^{\rm ecc.}_{\rm 3PN} &= -\frac{2355}{1462} e_{0}^{2} \nonumber \Big(\frac{v_{0}}{v}\Big)^{19/3}\Bigg[1+  \nonumber \bigg(\frac{ 299076223}{81976608} \\                     \nonumber
     &+ \frac{18766963}{2927736} \eta \bigg) v^{2} + \bigg(\frac{2833}{1008} - \frac{197}{36} \eta \bigg)v_{0}^{2} \\ 
     &- \frac{2819123}{282600} \pi v^{3} + \frac{377}{72}\pi v_{0}^{3} + \cdots + \mathcal{O}(v^{6})\Bigg]. 
\end{align}
We see that $\Delta\Psi^{\rm ecc.}_{\rm 3PN}$ is an expansion in powers of $v$ and $v_{0}$ in the small-eccentricity limit. Here $v_{0}$ = $[\pi M (1+z) f_{0}]^{1/3}$, and $f_{0}$ is the reference frequency at which the binary's instantaneous eccentricity $e_t$ equals $e_{0}$. See Eq.~(6.26) of \cite{Moore:2016qxz} for the full expression to 3PN order. Note that this waveform only models the inspiral phase of the binary evolution.

Since orbital eccentricity decays rapidly during the evolution of the binary, it is likely that a compact-object binary will have small eccentricity when it is observed by ground-based detectors. Hence, the {\tt TaylorF2Ecc} waveform is sufficient to quantify the effect of small eccentricity on parameter estimation. {\tt TaylorF2Ecc} is valid for small eccentricities  $e_0 \lesssim 0.2$ for comparable-mass systems and $e_0 \lesssim 0.02$ for extreme-mass ratio binaries ~\cite{Moore:2016qxz}.  

When considering alternative theories in the TGR framework, there will of course also be theory-dependent corrections (analogous to the $\delta \hat{\phi}_k$) to the PN coefficients in Eq.~\eqref{eccentric phase}. However, our purpose is not to generalize the TGR formalism to eccentric systems. Rather, we are considering the scenario where GR is correct and one is placing bounds on the TGR parameters given the assumption of a circular binary (as current LIGO/Virgo analyses in fact presume). In that case, an eccentric binary in GR will induce a bias in the TGR parameters. If that bias exceeds the statistical error bounds on the $\delta \hat{\phi}_k$ (which should have actual values of zero), one might falsely conclude that GR is violated.

\section{Statistical and Systematic Parameter error formalism}\label{section3}
We employ the Fisher information matrix formalism \cite{Cutler_Flanagan,Clifford_Will} for the estimation of parameters. The inner product between two signals $h_{1}(t)$ and $h_{2}(t)$ is given by
\begin{equation}\label{innerproduct}
    (h_{1}|h_{2}) = 2 \int_{0}^{\infty}\frac{\Tilde{h}_{1}^{*}(f) \Tilde{h}_{2}(f) + \Tilde{h}_{1}(f)\Tilde{h}_{2}^{*}(f) }{S_{n}(f)}\, df,
\end{equation}
where $\Tilde{h}(f)$ is the Fourier transform of $h(t)$, $S_{n}(f)$ is the one-sided noise power spectral density (PSD) of the detector, and $\ast$ denotes complex conjugation. The signal-to-noise ratio (SNR) $\rho$ is defined via the norm of the signal,
\begin{equation}\label{snr}
    \rho^{2} = (h|h) = 4\int_{0}^{\infty}\frac{|\Tilde{h}(f)|^{2}}{S_{n}(f)}\,df\,.
\end{equation}
In the limit of large SNR and stationary, Gaussian random noise, the probability that the signal $s(t)$ is characterized by the source parameters $\theta^{a}$ is given by 
\begin{equation}\label{posterior}
    p(\boldsymbol{\theta}|s)\; \propto \; p^{0}(\boldsymbol{\theta}) \exp\Big[-\frac{1}{2}\Gamma_{ab}(\theta^{a} - \hat{\theta}^{a})(\theta^{b} - \hat{\theta}^{b})\Big],
\end{equation}
where $p^{0}(\boldsymbol{\theta})$ is the prior probability based on our prior information about $\boldsymbol{\theta}$. In the absence of waveform modeling errors,  $\hat{\theta}^{a}$ are the true values of the source and TGR parameters (i.e., the parameter values where the probability distribution function is a maximum). The Fisher matrix $\Gamma_{ab}$ is given by the inner product 
\begin{equation}\label{fisher}
    \Gamma_{ab} = \Bigg(\frac{\partial h}{\partial \theta^{a}}\Bigg|\frac{\partial h}{\partial \theta^{b}}\Bigg)\,.
\end{equation}
If the prior probability $p^{0}(\boldsymbol{\theta})$ follows a Gaussian distribution peaked at $\theta^a = \bar{\theta}^a$,
\begin{equation}\label{prior}
 p^{0}(\boldsymbol{\theta}) \propto \exp\Big[-\frac{1}{2}\Gamma^{0}_{ab}(\theta^{a} - \Bar{\theta}^{a})(\theta^{b} - \Bar{\theta}^{b})\Big],
\end{equation}
then the covariance matrix is given by
\begin{equation}\label{covariance2}
     \Sigma_{ab} = (\Gamma_{ab}+ \Gamma^{0}_{ab})^{-1}  \,,
\end{equation}
where $\Gamma^{0}_{ab}$ denotes the Fisher matrix for the prior, and we assume $\bar{\theta}^{a} = \hat{\theta}^{a}$.
The 1$\sigma$ statistical errors  $\sigma_a$ in the parameters $\theta^{a}$ are given by the square root of the diagonal terms of the covariance matrix,
\begin{equation}\label{error}
    \sigma_{a} = \sqrt{\Sigma_{aa}}\,.
\end{equation}

In this study we consider the seven-dimensional parameter space:
\begin{equation}\label{parameter space}
    \theta^{a} = \{t_{c}, \phi_{c}, \ln M, \ln \eta, \chi_{1}, \chi_{2}, \delta\hat{\varphi}_{k}\},
\end{equation}
where $\delta\hat{\varphi}_{k}$ represents any {\it one} of the TGR parameters. (We drop the ``log'' label here and below for convenience.)  By computing the Fisher matrix and applying Eq.~\eqref{error}, we compute the $1\sigma$ error bounds. This is done separately for each of the nine possible choices for $\delta\hat{\varphi}_{k}$. Varying more than one $\delta\hat{\varphi}_{k}$ coefficient at a time leads to uninformative bounds on $\delta\hat{\varphi}_{k}$ due to correlations among the different parameters \cite{Arun:2006yw,Gupta:2020lxa,Datta:2020vcj}. In practice, a true GR deviation would likely modify multiple $\delta\hat{\varphi}_{k}$ coefficients simultaneously. However, the approach applied here can pick up those deviations effectively \cite{Sampson:2013lpa,Meidam:2017dgf,Haster:2020nxf}. Recognizing that the coalescence phase and spin parameters are physically restricted to the ranges $\phi_{c} \in [-\pi, \pi]$ and $\chi_{1,2} \in [-1, 1]$, we attempt to incorporate this constraint in the Fisher matrix approach by imposing Gaussian priors on those parameters with zero means and $1\sigma$ widths given by $\delta \phi_c =\pi$ and $\delta \chi_{1,2}=1$ \cite{Favata:2021vhw}. These priors improve the condition number of the Fisher matrix. No priors are placed on the TGR coefficients or other parameters. 

To calculate the systematic parameter biases due to unmodeled binary eccentricity, we apply the formalism developed by Cutler and Vallisneri \cite{Cutler:2007mi}.
Consider a GW signal described by a ``true'' (exact) waveform model $\Tilde{h}_{\rm T}(\theta_{\rm T})$ that depends on the true system parameters $\theta_{\rm T}^{a}$. This signal is analyzed using an approximate waveform model $\Tilde{h}_{\rm AP}(\theta_{\rm bf})$ that depends on the ``best fit'' model parameters $\theta_{\rm bf}^a$. (Under our Gaussian model assumption, $\hat{\theta}^a \rightarrow \theta_{\rm bf}^a$, which becomes the peak of our recovered Gaussian distribution and is offset from the true value $\theta_{\rm T}^a$.) We can express the approximate and true waveforms in terms of an amplitude ${\mathcal A}$ and phase $\Psi$ via
\begin{equation}\label{approximate waveform}
    \Tilde{h}_{\rm AP}(\theta_{\rm bf}) = \mathcal{A}_{\rm AP}(\theta_{\rm bf})  e^{i \Psi_{\rm AP}(\theta_{\rm bf})} \;\;\;\; \text{and}
\end{equation}
\begin{align}
\label{true waveform}
    \Tilde{h}_{\rm T}(\theta_{\rm T}) &= \mathcal{A}_{\rm T}(\theta_{\rm T})  e^{i \Psi_{T}(\theta_{T})} \,, \\
 &= \left[\mathcal{A}_{\rm AP}(\theta_{\rm bf}) + \Delta\mathcal{A} \right]  e^{i [\Psi_{\rm AP}(\theta_{\rm bf}) + \Delta\Psi]} \;, \nonumber 
\end{align}
where $\Delta\mathcal{A}$ and $\Delta\Psi$ in the last line represent the difference between the true (T) and approximate (AP) waveform amplitudes and phases. If we define
\begin{equation}\label{systemtic defintion}
   \Delta \theta^{a} = \theta_{\rm T}^a - \theta_{\rm bf}^a,
\end{equation}
as the systematic error in the parameter $\theta^{a}$, then 
$\Delta \theta^{a}$ can be approximated by \cite{Cutler:2007mi}
\begin{equation}\label{systematic errors}
    \Delta \theta^{a} \approx \Sigma^{ab}({\theta_{\rm bf}})\bigg(\big[\Delta\mathcal{A} + i \mathcal{A}_{\rm AP} \Delta\Psi \big] e^{i \Psi_{\rm AP}} \bigg| \partial_{b} \Tilde{h}_{\rm AP}(\theta_{\rm bf})\bigg) \,,
\end{equation}
where $\Sigma^{ab}$ is the covariance matrix calculated using the approximate waveform. Note that we have dropped the parameter index on $\theta^a$ for simplicity when appearing inside a function argument. See \cite{Favata:2021vhw} for more details.

Since $\Sigma^{ab}$ scales as $\rho^{-2}$ to leading order and the inner product on the right-hand side of Eq.~\eqref{systematic errors} is proportional to $\rho^{2}$, $\Delta \theta^{a}$ is independent of $\rho$. In contrast, the statistical error scales like $\rho^{-1}$ in the limit of high SNR. This suggests that systematic errors could exceed statistical errors in the high SNR limit (when statistical errors become small). Hence, small deviations between the approximate and true waveforms can bias the estimated parameter values.

In our case, we are interested in the bias resulting from a neglect of binary eccentricity. To model this, we apply the same approach as in \cite{Favata:2021vhw}, setting $\Delta {\mathcal A}$ to zero and considering a waveform deviation $\Delta \Psi$ in the SPA phase that arises from the eccentric corrections contained in $\Delta \Psi_{\rm 3PN}^{\rm ecc.}$ [Eqs.~\eqref{decomposed phase} and \eqref{eccentric phase}]. The approximate waveform is taken to be $\Psi_{\rm AP} = \Psi(f)^{\rm circ., TGR}$.

The above formalism is applied to the LIGO \cite{LIGOScientific:2014pky} and Cosmic Explorer (CE) \cite{CE:2019iox} detectors. We use the LIGO noise PSD from Eq.~(4.7) of \cite{Ajith:2011ec}. For CE, we use the PSD from Eq.~(3.7) of \cite{Kastha:2018bcr}. When evaluating the integrals in Eq.~\eqref{innerproduct} and \eqref{snr}, the lower frequency of integration is chosen to be $10$ Hz for LIGO and $5$ Hz for CE (corresponding to the detectors' low-frequency sensitivity limits). Since we use inspiral-dominated waveforms, when considering BBHs the upper limit of the integrals is chosen as the innermost stable circular orbit $(f_{\rm isco})$ of the remnant BH \cite{1972ApJ...178..347B,Husa:2015iqa,Hofmann:2016yih}. This frequency is a function of the two component masses ($m_{1}$ and $m_{2}$), their dimensionless spins  ($\chi_{1}$  and $\chi_{2}$), and the source redshift $z$. See Appendix C of \cite{Favata:2021vhw} for the full expression. When considering BNS, we choose the upper-frequency cutoff to be $1000\,{\rm Hz}$ for LIGO and $1500\,{\rm Hz}$ for CE. The reference frequency $f_0$ is chosen to be $10$ Hz for both LIGO and CE.

\section{\label{section4}Results}
Our main goal is to quantify the systematic bias on the TGR parameters $\delta \hat{\varphi}_k$ (which are zero in GR) due to the neglect of orbital eccentricity in the waveform model. In particular, we are interested in the eccentricity value where the systematic errors are equal to the statistical errors on the TGR parameters. When systematic bias exceeds the statistical errors, the use of eccentric waveforms becomes essential for the accurate estimation of the source parameters and for tests of GR. We consider four representative binary black hole (BBH) systems with total masses $(15 M_{\odot}, 35 M_{\odot}, 65 M_{\odot}, 90 M_{\odot})$. All have mass ratio $2:1$ and are located at a luminosity distance $d_{L}=500\, {\rm Mpc}$. These systems have SNRs of $11.80$ $(413.92)$, $23.41$ $(834.06)$, $37.41$ $(1380.95)$, $46.74$ $(1790.35)$ in the LIGO (CE) bands, respectively. The spins are assumed to be aligned with the orbital angular momentum of the binary, with dimensionless spin magnitudes $\chi_{1}=0.5$ and $\chi_{2}=0.4$. We also consider a binary neutron star (BNS) system with component masses $m_1=1.4 M_{\odot}$, $m_2=1.2 M_{\odot}$, and component spins $\chi_1=\chi_2=0.05$ at a distance of $100\, \text{Mpc}$. This system has an SNR of $13.61$ ($476.62$) in the LIGO (CE) band. These values are representative of ``typical'' binaries that might yield strong constraints on the TGR parameters. At the end of this section we briefly discuss how other mass ratios and spins affect our results.

\begin{figure*}[ph]
    \centering 
    \begin{subfigure}{\includegraphics[width=0.42\textwidth]{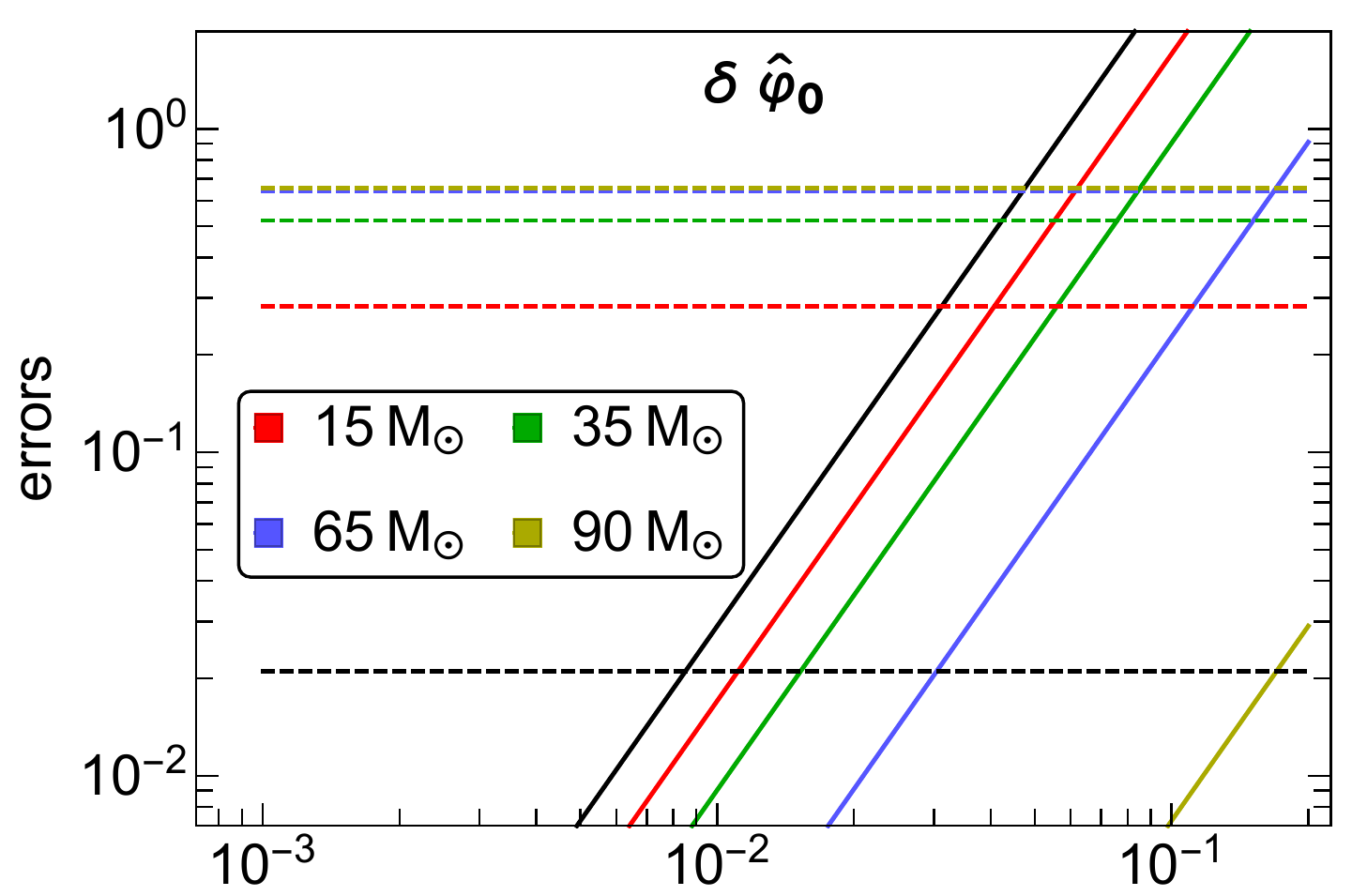}}
    \end{subfigure}
    \vspace{-0.35cm}
    \begin{subfigure}{\includegraphics[width=0.42\textwidth]{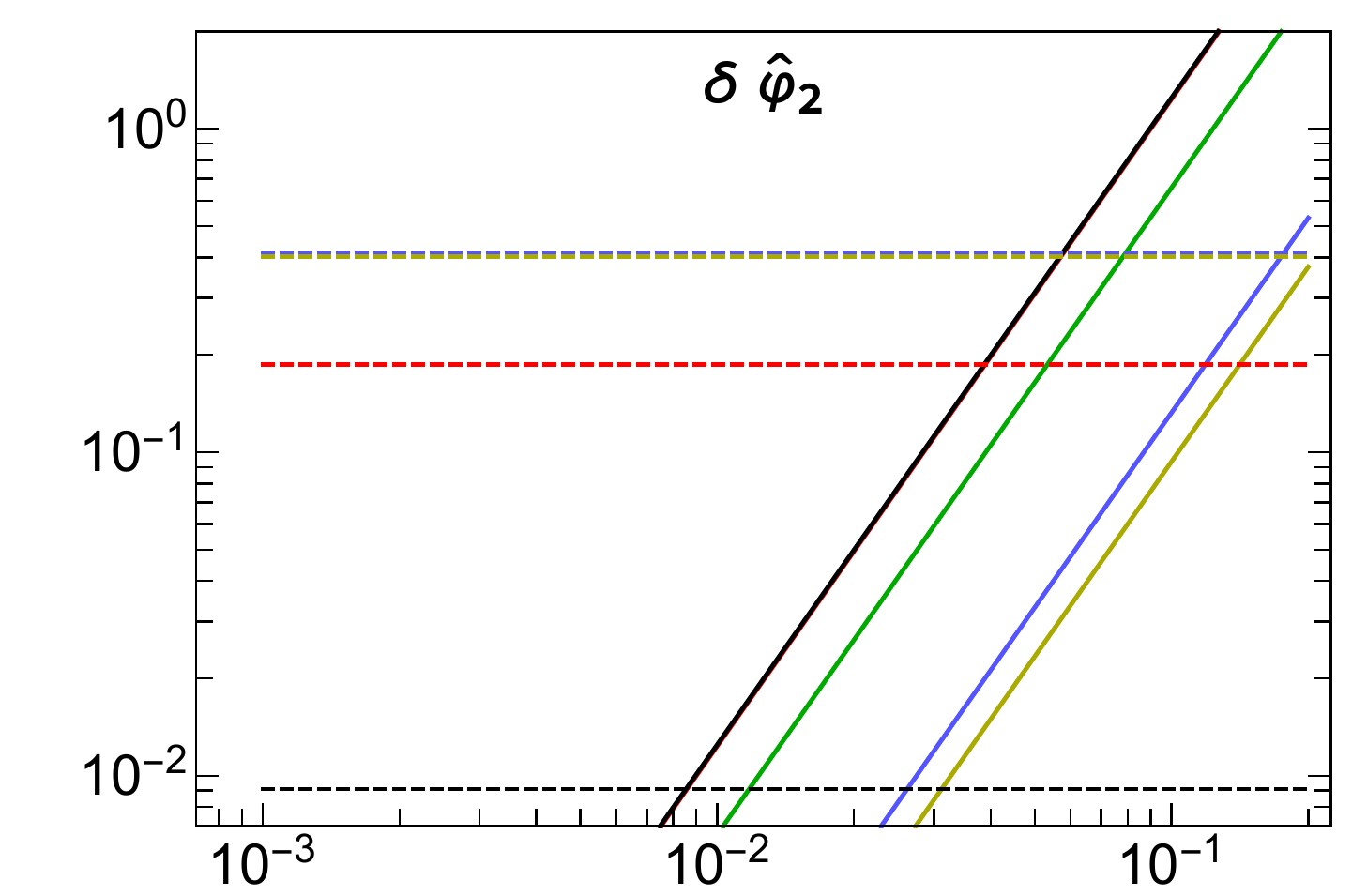}} 
     \end{subfigure}
     \begin{subfigure}{\includegraphics[width=0.42\textwidth]{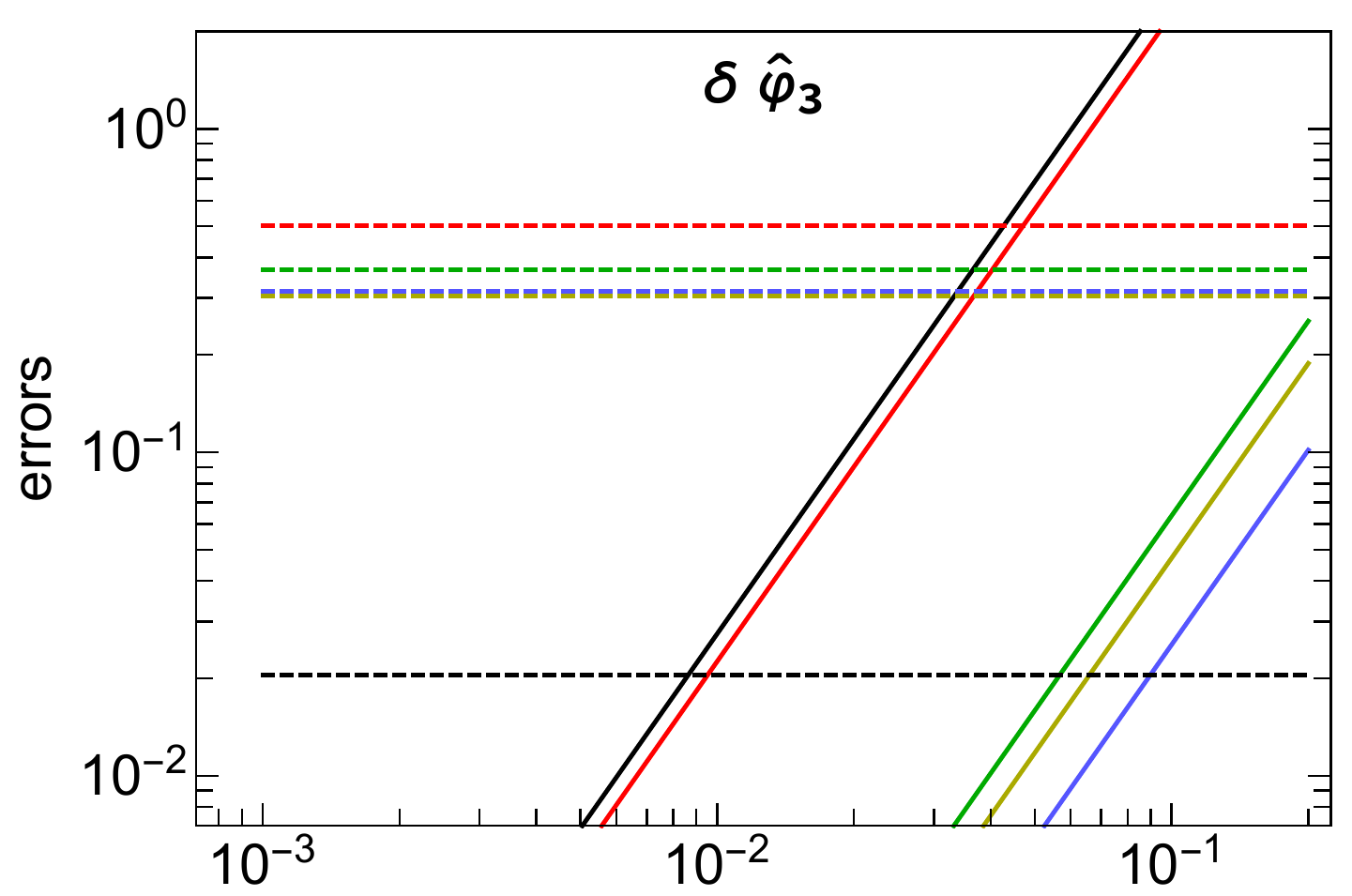}} 
     \end{subfigure}
      \vspace{-0.35cm}
    \begin{subfigure}{\includegraphics[width=0.42\textwidth]{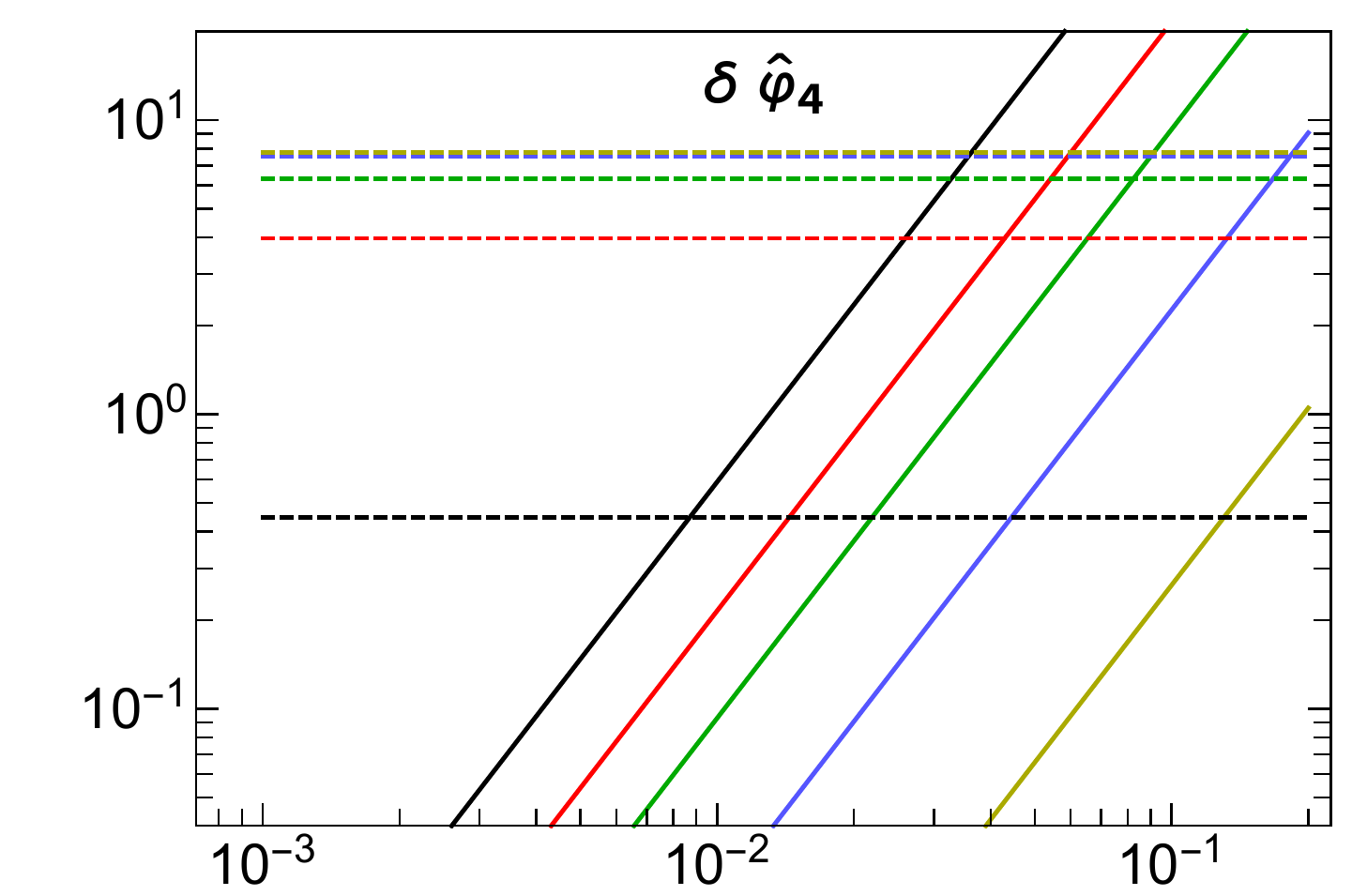}} 
     \end{subfigure}
      \begin{subfigure}{\includegraphics[width=0.42\textwidth]{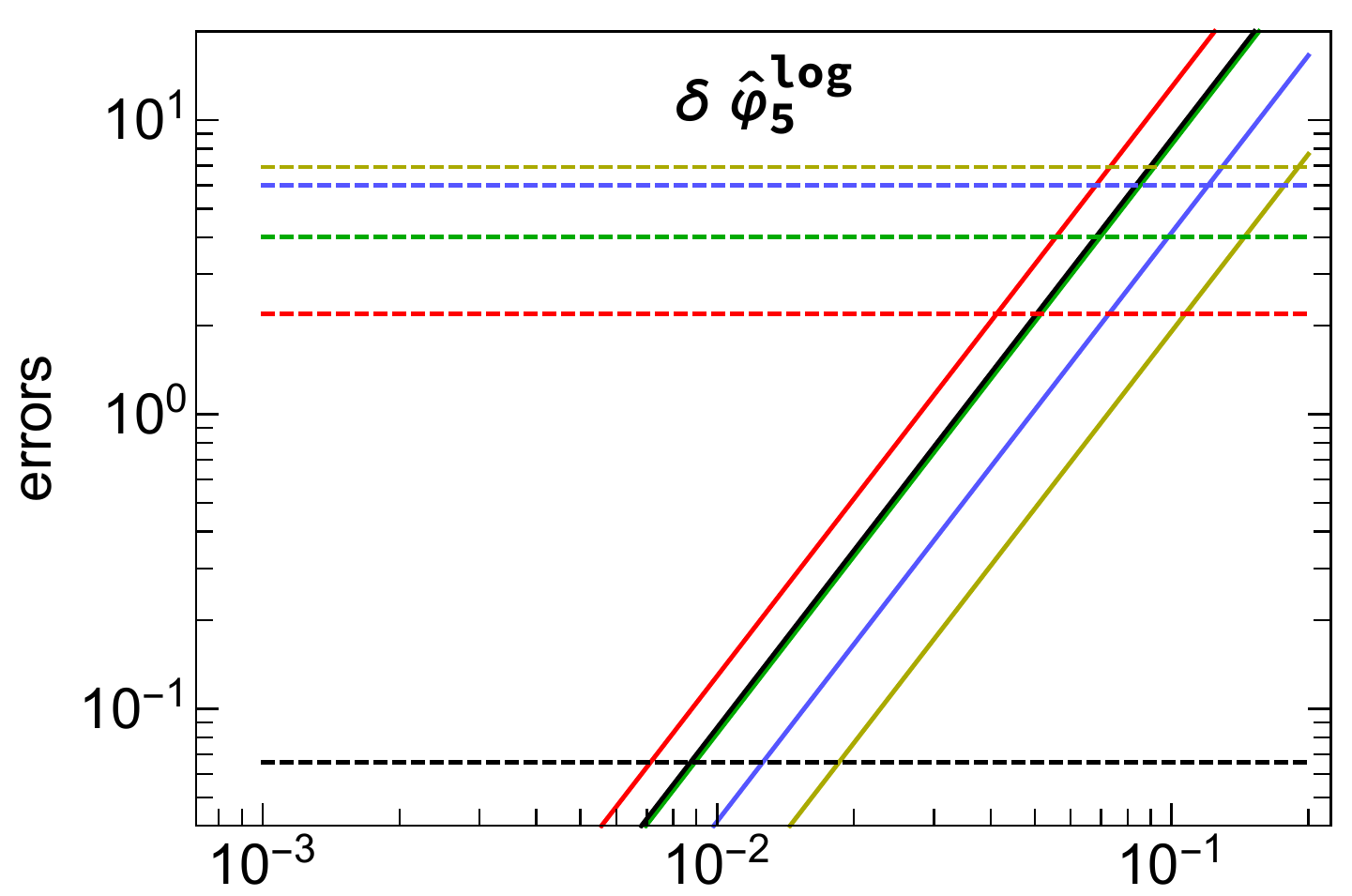}} 
     \end{subfigure}
      \vspace{-0.35cm}
    \begin{subfigure}{\includegraphics[width=0.42\textwidth]{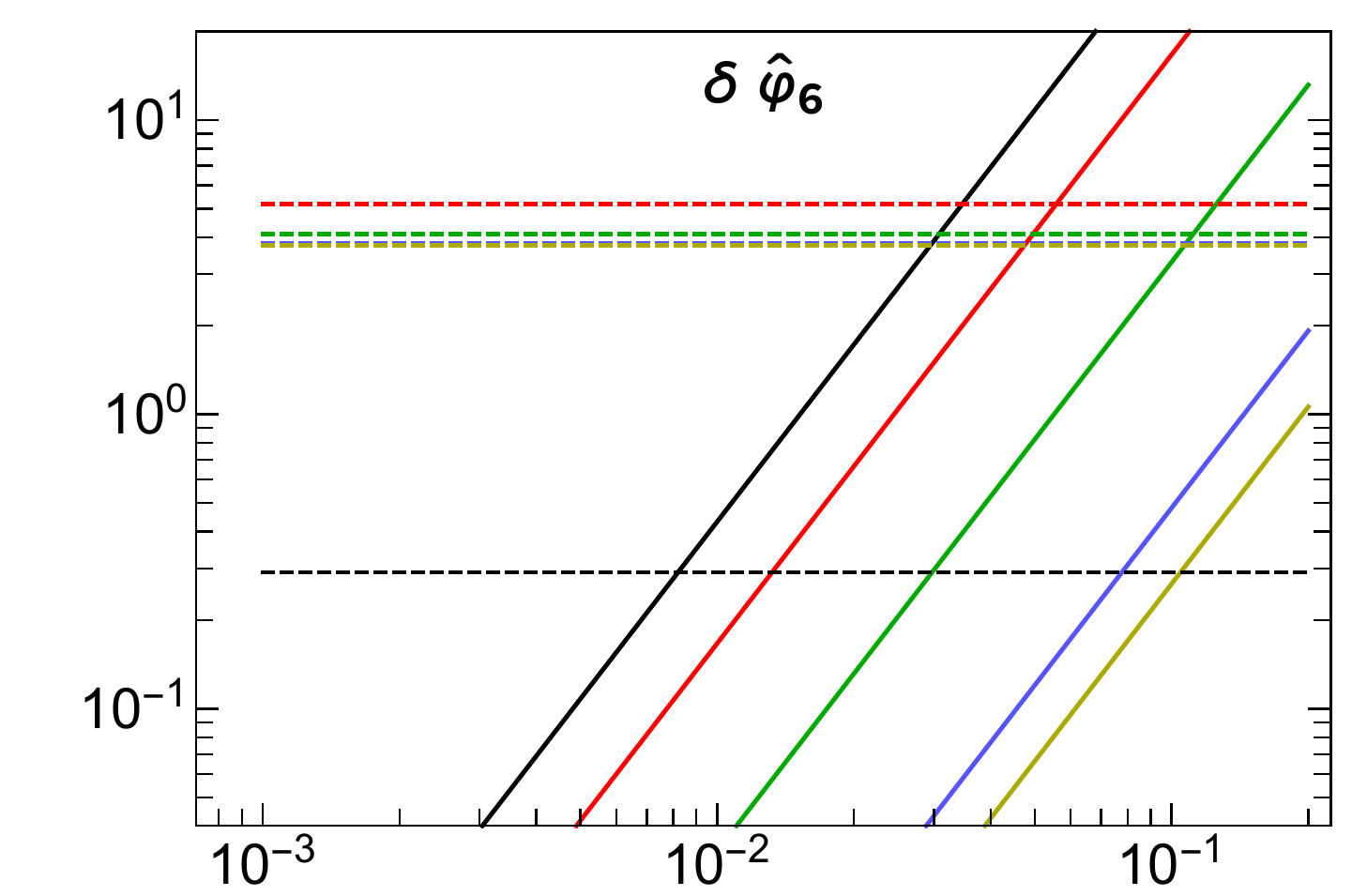}} 
     \end{subfigure}
      \begin{subfigure}{\includegraphics[width=0.42\textwidth]{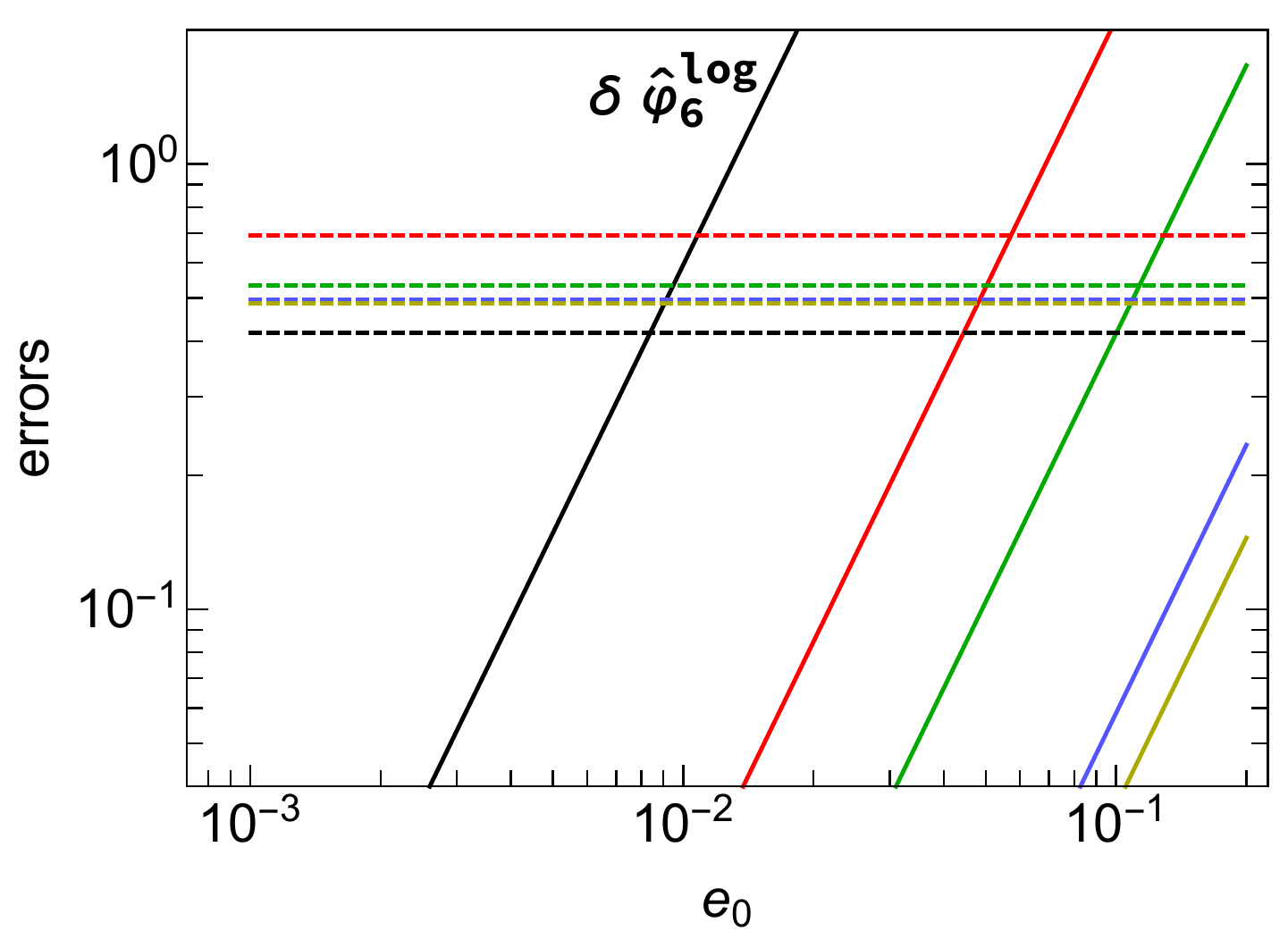}} 
     \end{subfigure}
     \begin{subfigure}{\includegraphics[width=0.42\textwidth]{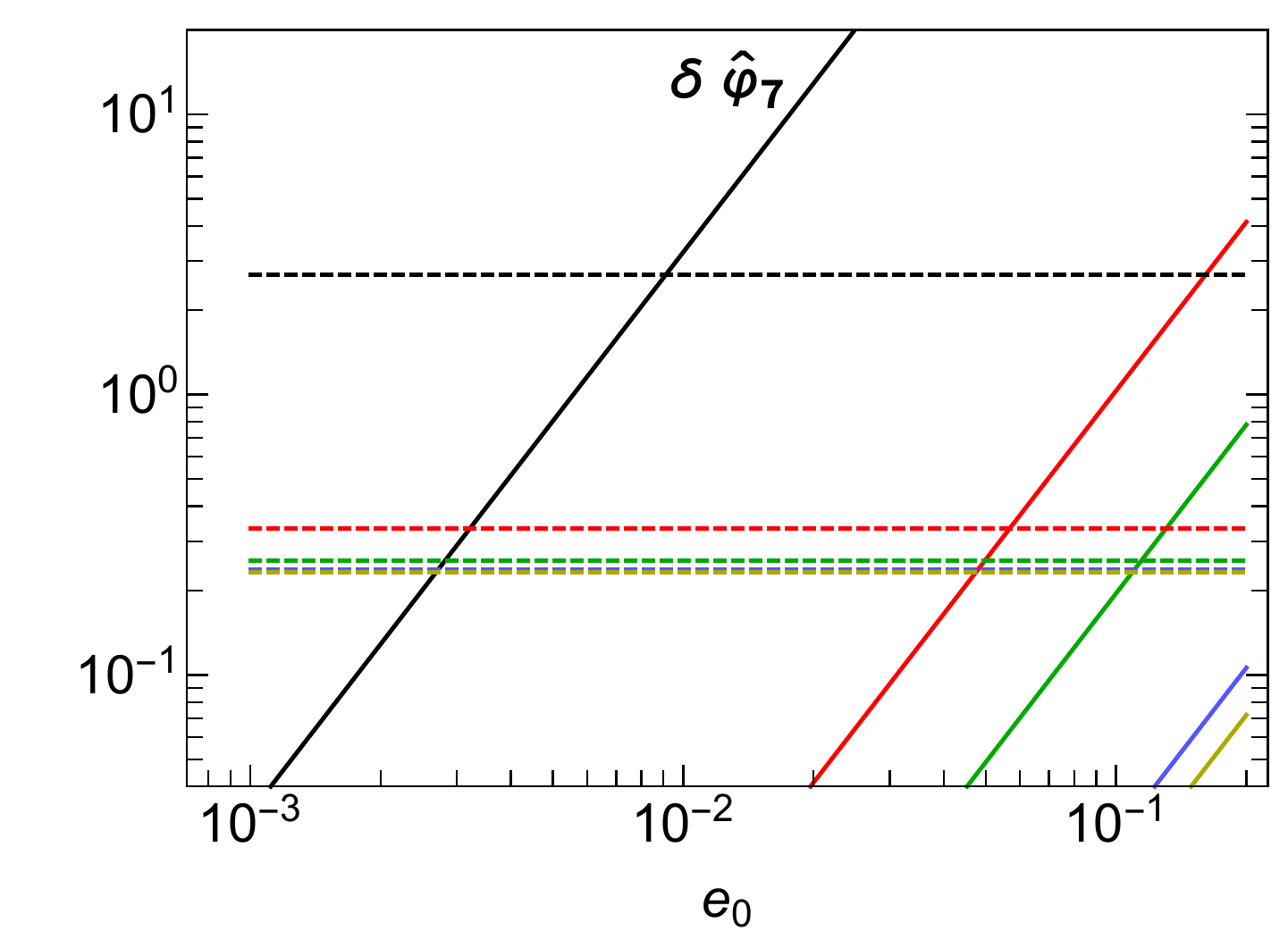}} 
      \end{subfigure}
    \caption{(Color online) Systematic and statistical errors on the TGR parameters $\delta\hat{\varphi}_{k}$ as a function of the binary eccentricity parameter $e_0$ for various sources in the LIGO band. The eccentricity $e_0$ is defined at $10$ Hz. Systematic errors are shown by slanted solid lines. (Note that these are the absolute values of the systematic errors.) The $1\sigma$ statistical errors are shown by horizontal dashed lines. The different colors indicate sources with different values of total mass $M$ (see legend). For all sources the mass ratio is fixed to $2:1$, the spin parameters are $\chi_{1}=0.5$, $\chi_{2}=0.4$, and the luminosity distance is $500$ Mpc. The black lines represent the systematic and statistical errors for a BNS system with component masses $m_1=1.4 M_{\odot}$, $m_2=1.2 M_{\odot}$ and component spins $\chi_1=\chi_2=0.05$ at a distance of $100$ Mpc.}
   \label{errors_LIGO}
\end{figure*}

\begin{figure*}[ph]
    \centering 
    \begin{subfigure}{\includegraphics[width=0.42\textwidth]{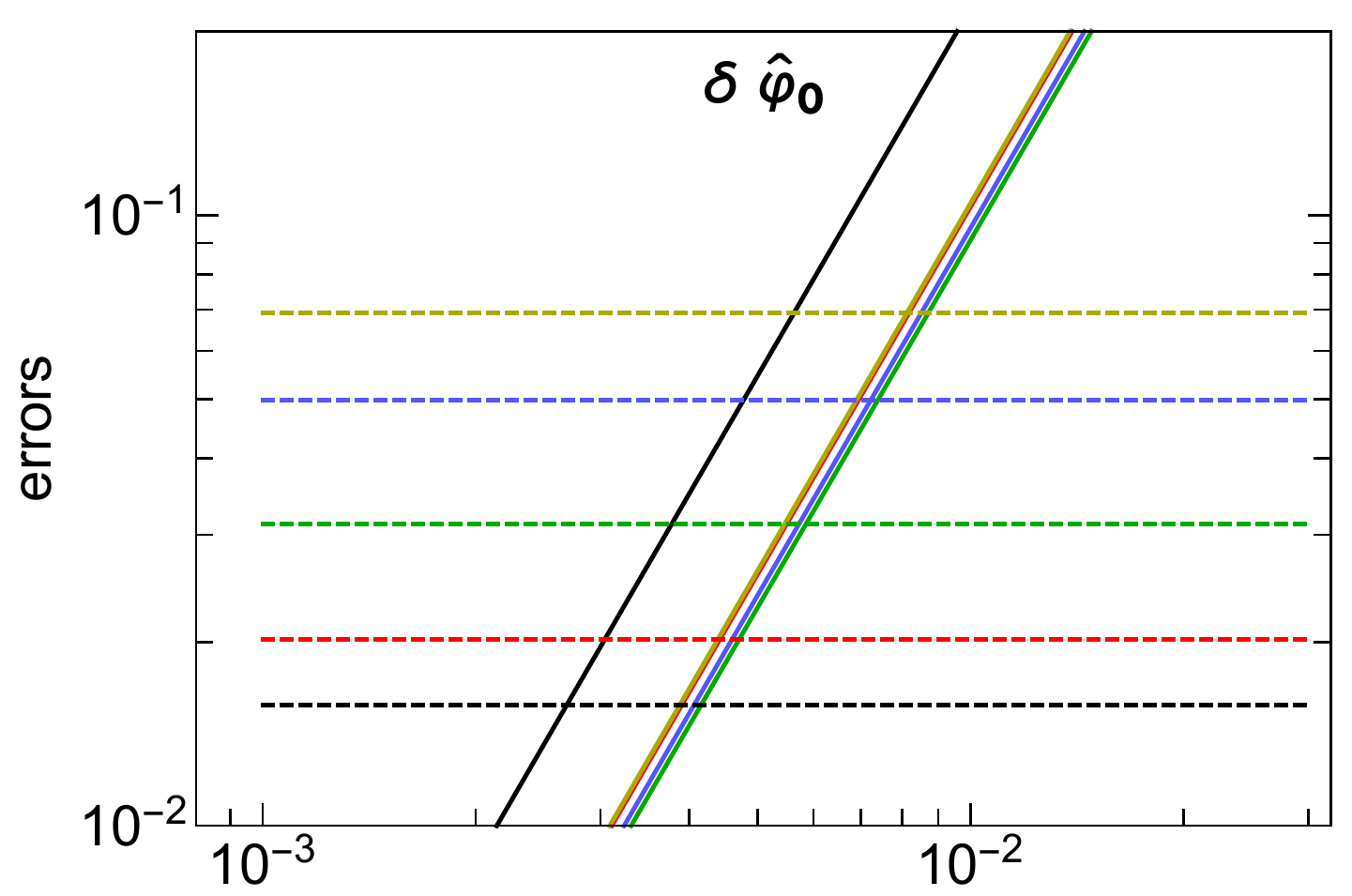}}
    \end{subfigure}
    \vspace{-0.35cm}
    \begin{subfigure}{\includegraphics[width=0.42\textwidth]{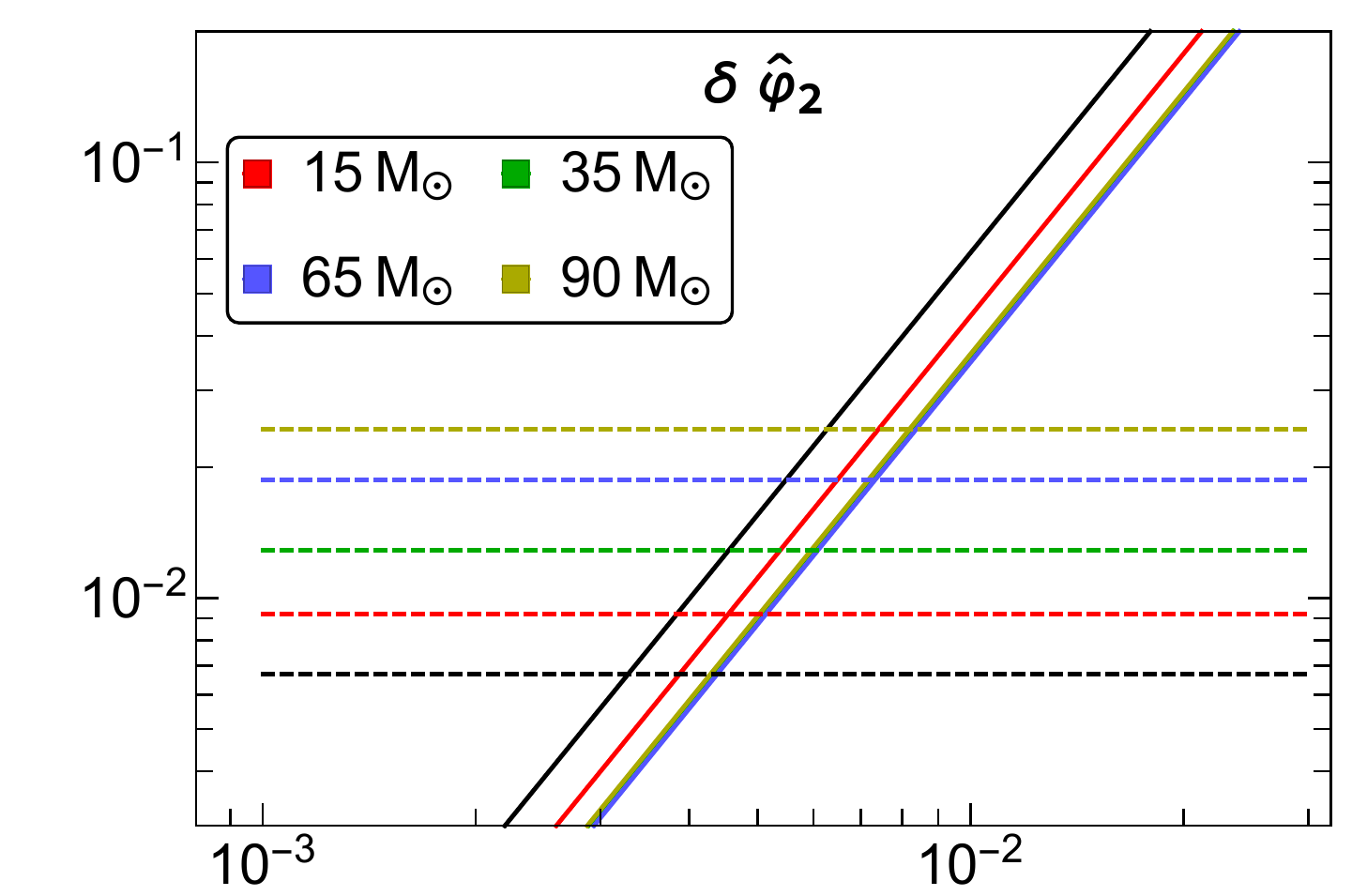}} 
     \end{subfigure}
     \begin{subfigure}{\includegraphics[width=0.42\textwidth]{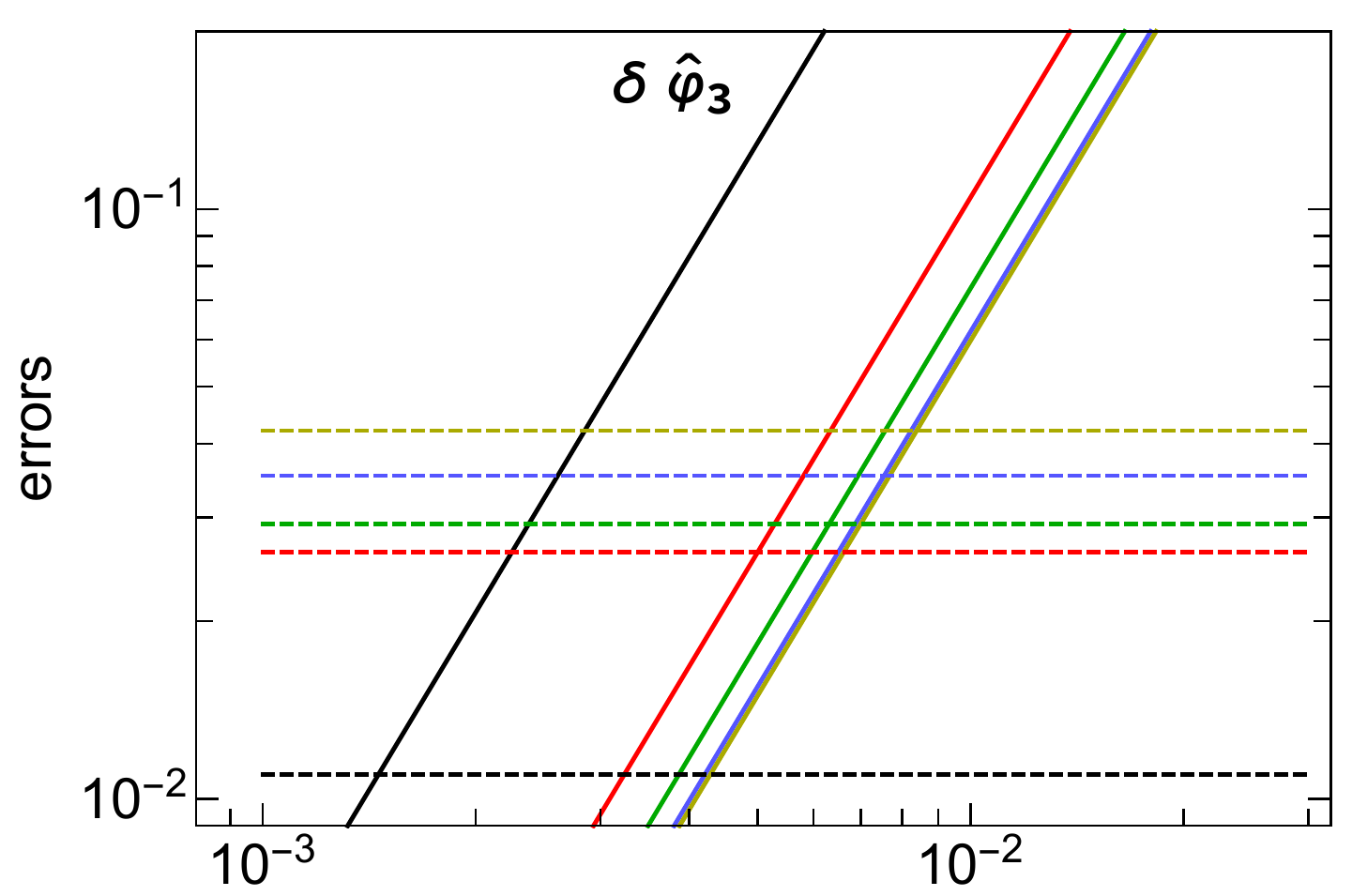}} 
     \end{subfigure}
      \vspace{-0.35cm}
    \begin{subfigure}{\includegraphics[width=0.42\textwidth]{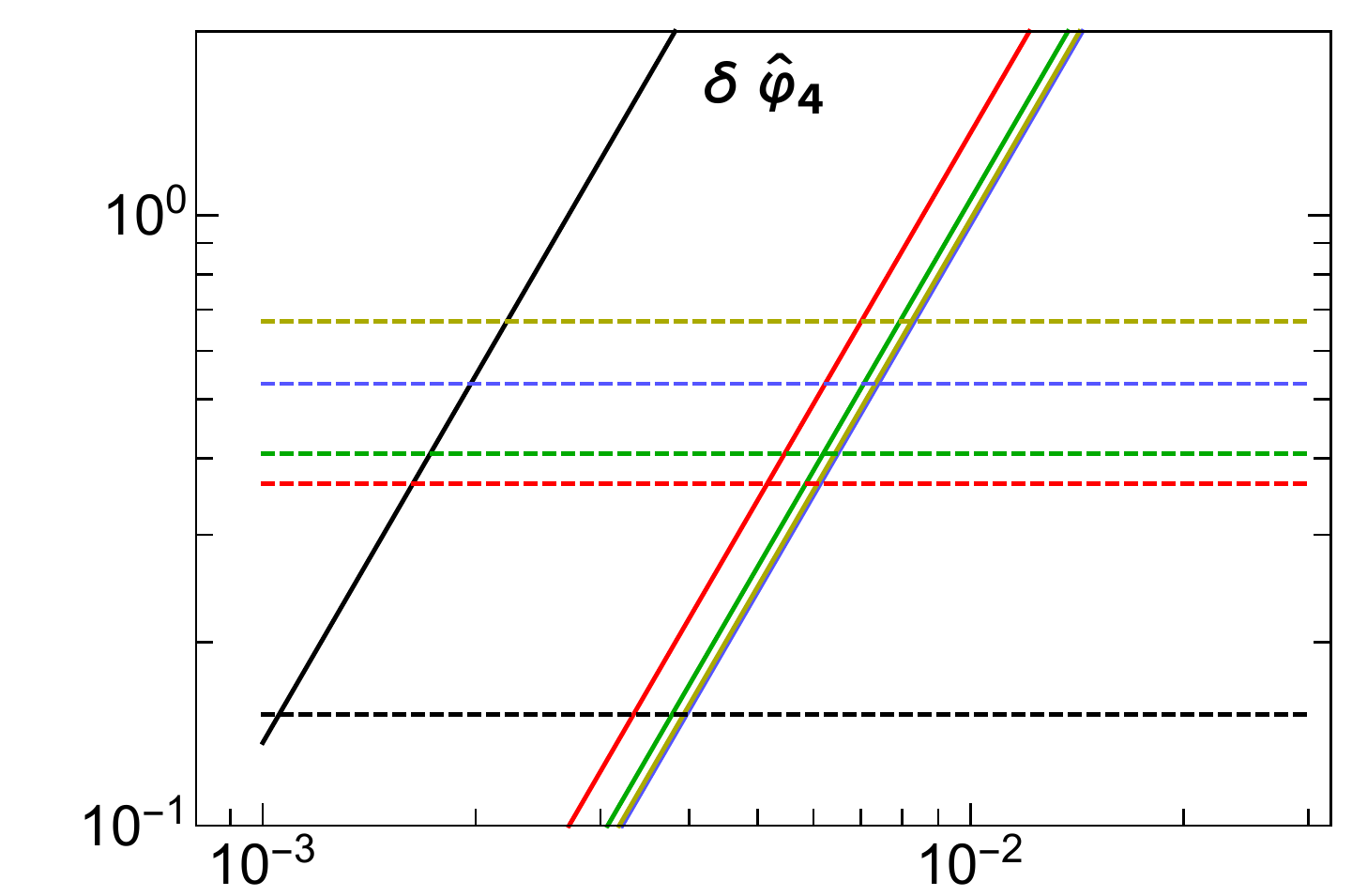}} 
     \end{subfigure}
      \begin{subfigure}{\includegraphics[width=0.42\textwidth]{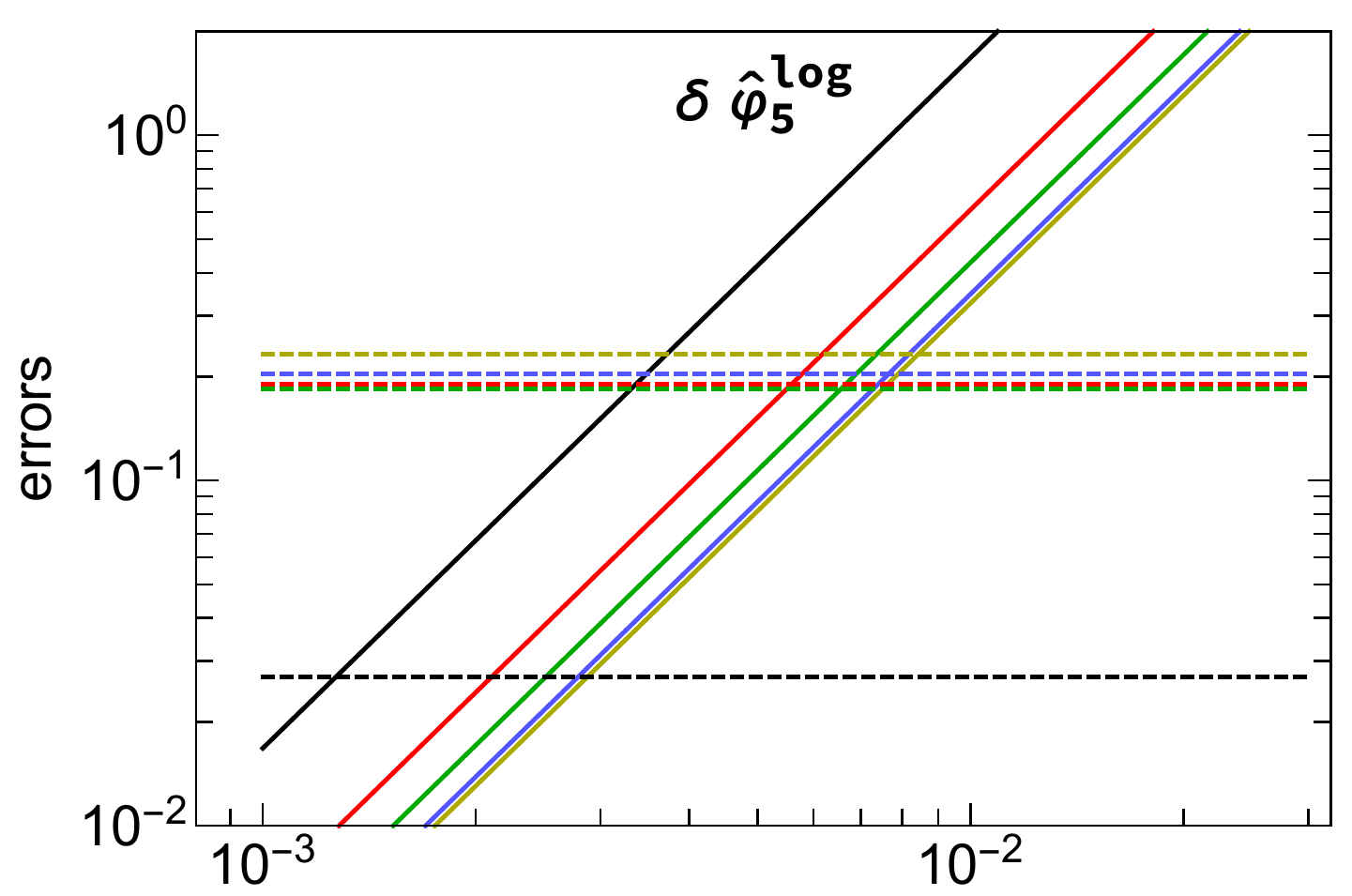}} 
     \end{subfigure}
      \vspace{-0.35cm}
    \begin{subfigure}{\includegraphics[width=0.42\textwidth]{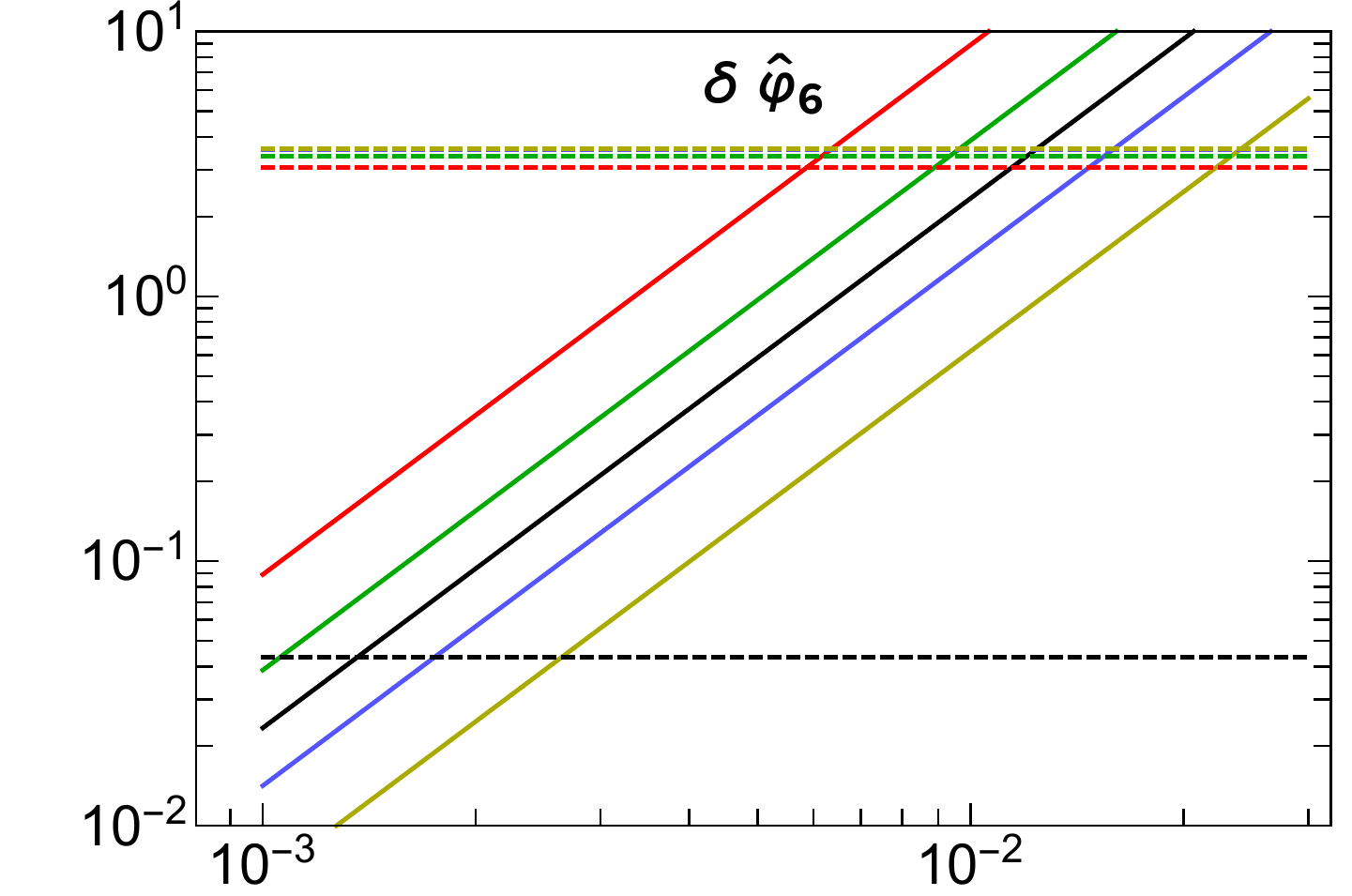}} 
     \end{subfigure}
      \begin{subfigure}{\includegraphics[width=0.42\textwidth]{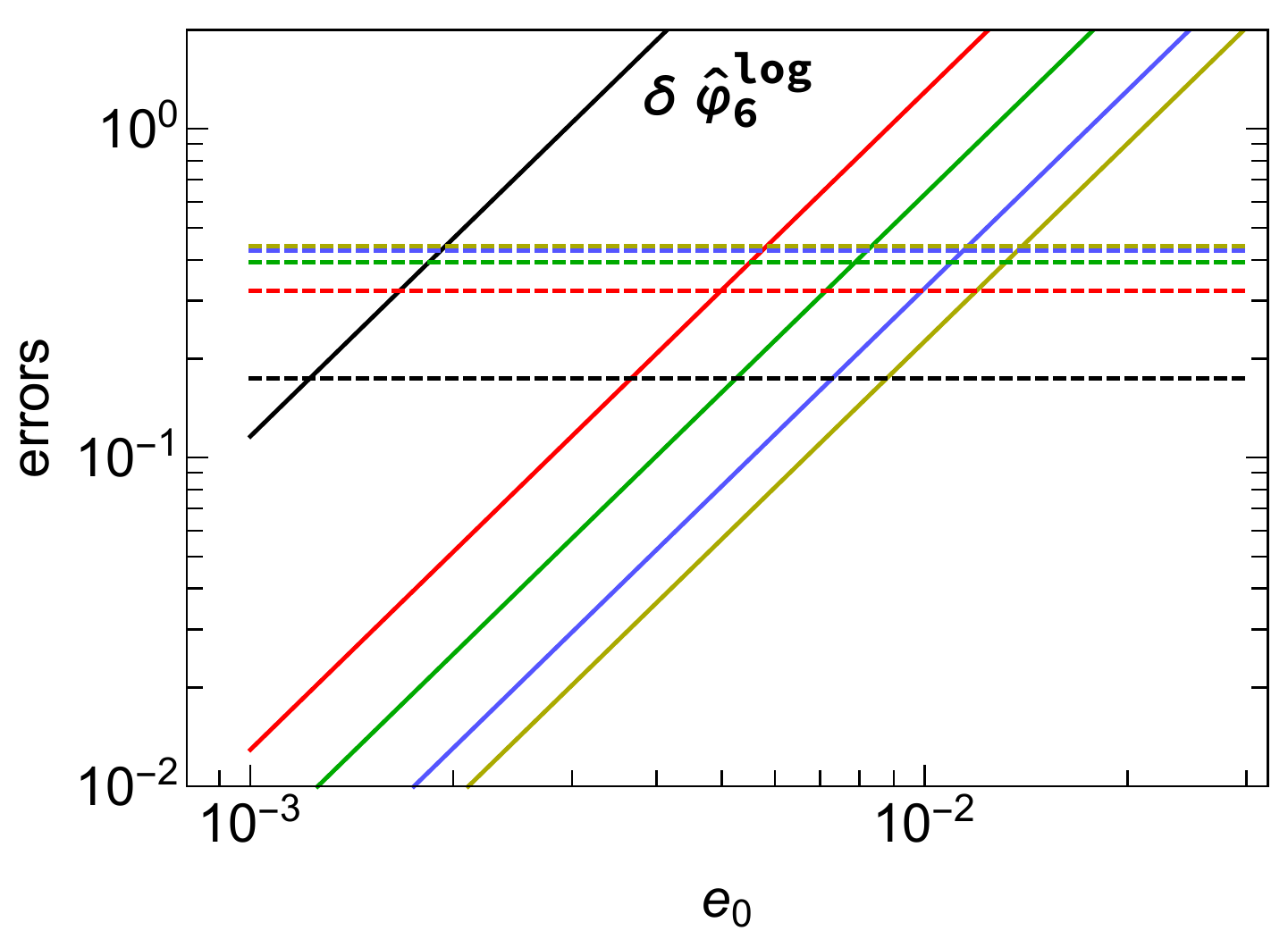}} 
     \end{subfigure}
     \begin{subfigure}{\includegraphics[width=0.42\textwidth]{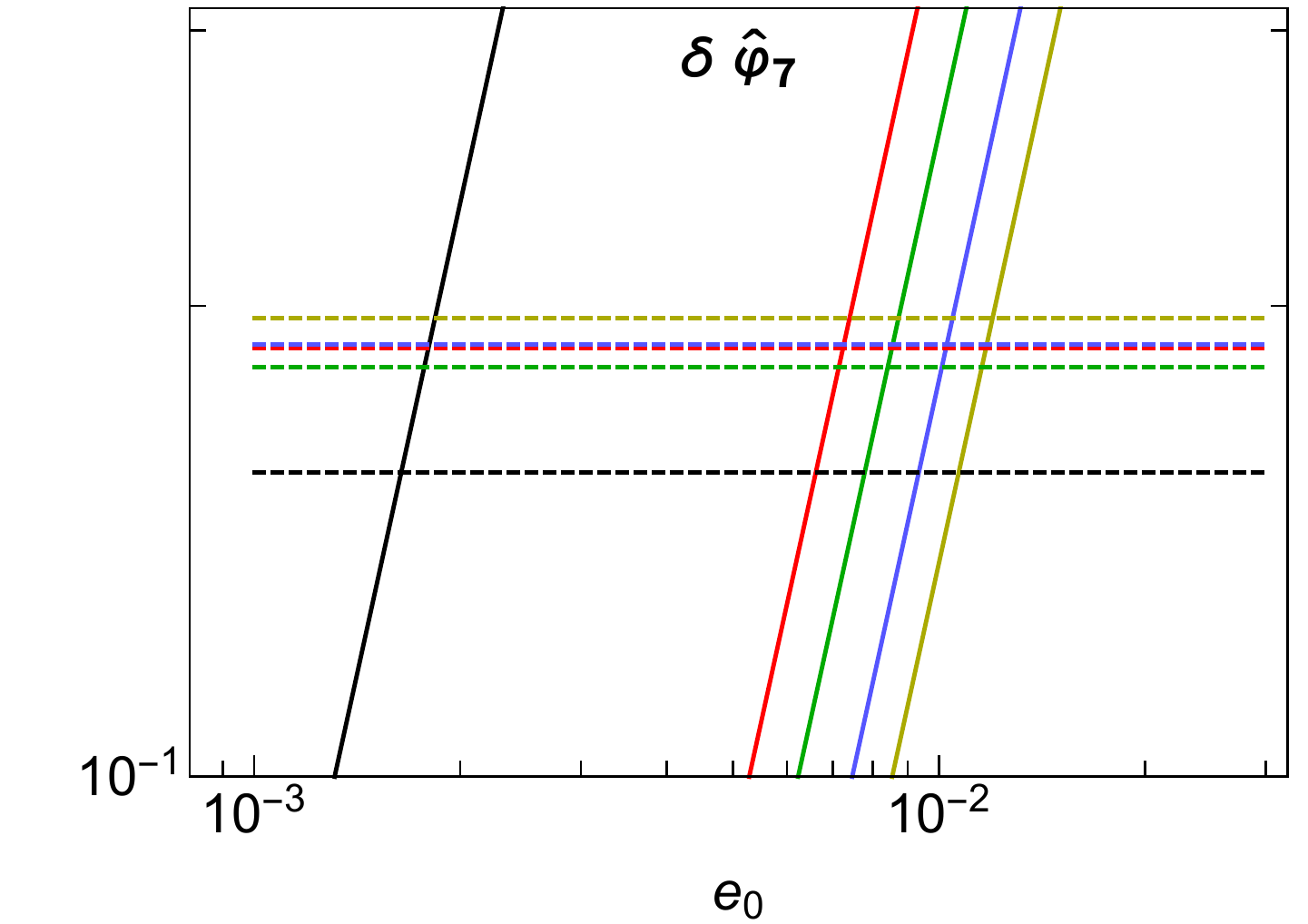}} 
      \end{subfigure}
     \caption{(Color online) Same as Fig. \ref{errors_LIGO} but for the Cosmic Explorer (CE) detector. The eccentricity parameter is again defined at $f_0=10$ Hz. The luminosity distance and all other parameters are identical to the choices in Fig. \ref{errors_LIGO}.}
   \label{errors_CE}
\end{figure*}

\begin{figure}[th]
    \centering 
    \begin{subfigure}{\includegraphics[width=0.48\textwidth]{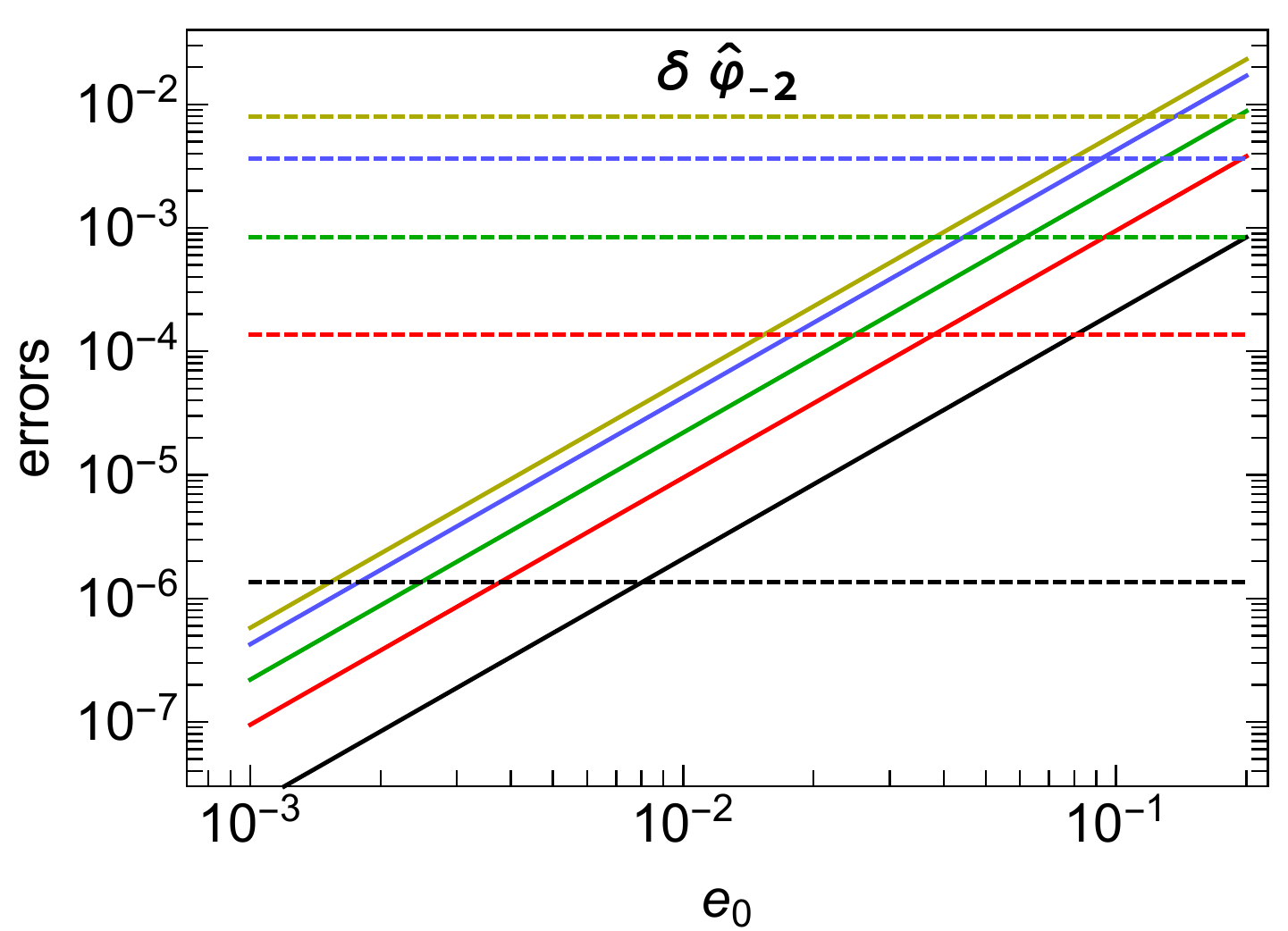}} 
    \end{subfigure}
  
  \begin{subfigure}{\includegraphics[width=0.48\textwidth]{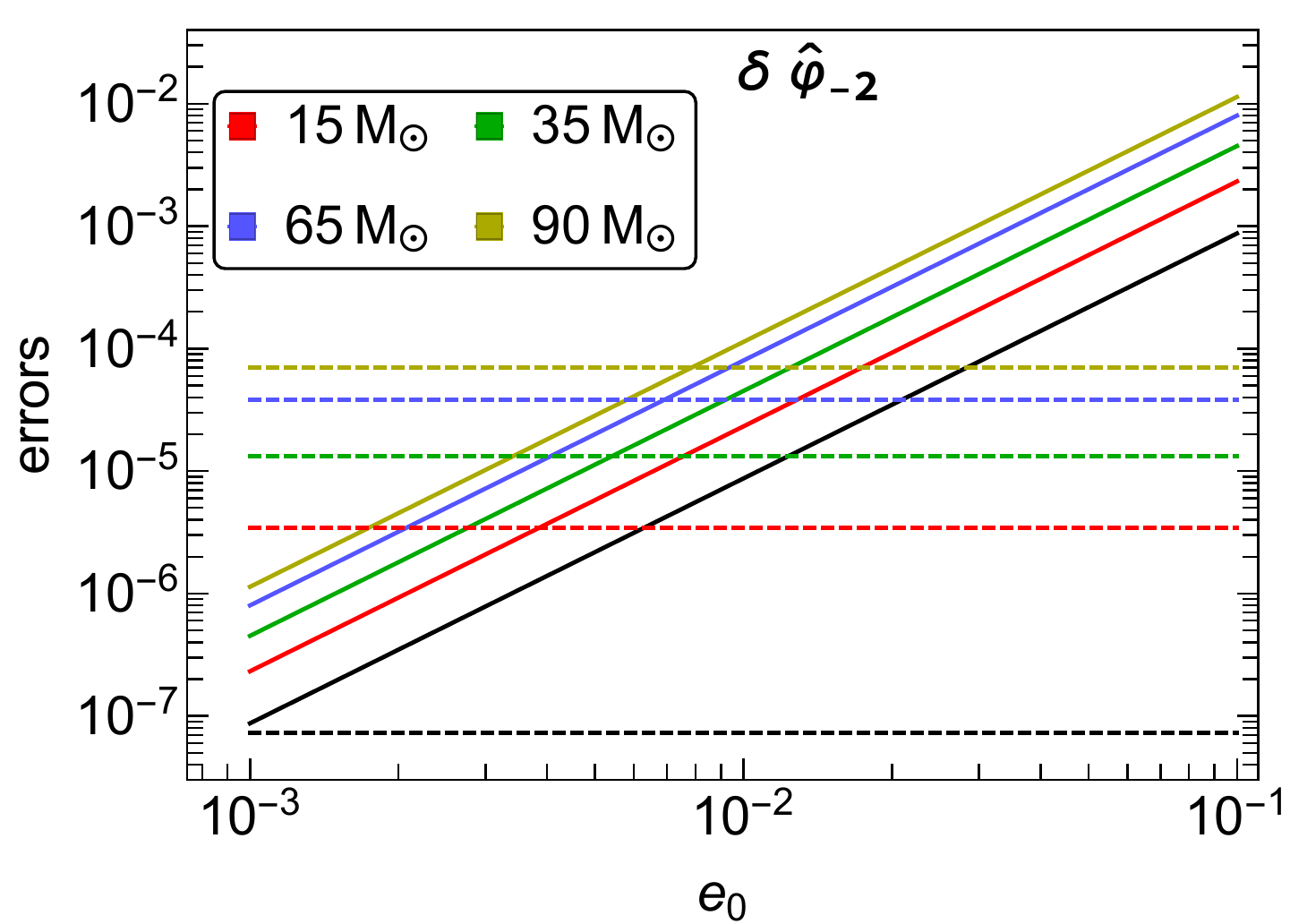}} 
    \end{subfigure}
     \caption{(Color online) Systematic and statistical errors for the dipole TGR parameter $\delta \hat{\varphi}_{-2}$. The top panel shows results for LIGO; the bottom for CE. The system parameters and labeling scheme are the same as in Figs.~\ref{errors_LIGO} and \ref{errors_CE}. In contrast to those figures, higher mass binaries show a larger systematic error at a given value of $e_0$. This is explained in the main text.}
   \label{errors_dipole}
\end{figure}

The main results of our analysis are shown in~\Cref{errors_LIGO,errors_CE,errors_dipole}. Figure~\ref{errors_LIGO} shows systematic errors (sloped, solid lines) and statistical errors (horizontal, dashed lines) for the various $\delta\hat{\varphi}_{k}$ (except $\delta \hat{\varphi}_{-2}$) as a function of eccentricity $e_0$ in the LIGO band. Systematic bias increases as the eccentricity of the source increases. (Recall that $e_0$ is defined at a reference frequency of $10$ Hz). Figure \ref{errors_CE} is the same, but uses the CE sensitivity curve. The statistical and systematic errors on the dipole parameter $\delta \hat{\varphi}_{-2} = \varphi_{-2}$ are shown separately (for LIGO and CE) in Fig. \ref{errors_dipole}. 



Recall that these TGR parameters [cf.~Eq.~\eqref{deformation}] are defined as the fractional deviation from the GR values of the PN coefficients in the waveform phasing.
To better understand the meaning of the errors presented in the figures, let us presume that GR is \emph{not} correct and the true (T) values of the PN coefficients are given by a nonzero $\delta \hat{\varphi}_k$, $\varphi_k^{\rm T}=\varphi_k^{\rm GR} (1+\delta\hat{\varphi}_k^{\rm T})$. A GW measurement then supplies a best-fit value of the TGR parameter $\delta \hat{\varphi}_k$ (offset from the true value by the systematic error) and its $1\sigma$ statistical error, $\delta \hat{\varphi}_k^{\rm bf} \pm \sigma_{\delta \hat{\varphi}_k} = \delta \hat{\varphi}_k^{\rm T} + \Delta(\delta \hat{\varphi}_k) \pm \sigma_{\delta \hat{\varphi}_k}$. One could equivalently supply the best-fit value of the PN coefficients and their statistical errors, $\varphi_k^{\rm bf} \pm \sigma_{\varphi_k} = \varphi_k^{\rm T} + \Delta \varphi_k \pm \sigma_{\varphi_k}$. The resulting systematic and statistical errors are related by
\begin{align}
\label{eq:Deltaphi}
    \Delta(\delta\hat{\varphi}_k) &= \delta\hat{\varphi}_k^{\rm bf} - \delta\hat{\varphi}_k^{\rm T} = \frac{\Delta \varphi_k}{\varphi_k^{\rm GR}} = \frac{\varphi_k^{\rm bf}-\varphi_k^{\rm T}}{\varphi_k^{\rm GR}} \,, \\
    \label{eq:deltaphi}
    \sigma_{\delta\hat{\varphi}_k} &= \frac{\sigma_{\varphi_k}}{\varphi_{k}^{\rm GR}} \,.
\end{align}
The left-hand sides of the above equations are the systematic and statistical errors supplied in Figs.~\ref{errors_LIGO} and \ref{errors_CE}. 

Note that the statistical error provided is equivalent to a fractional error in the PN coefficient relative to its GR value (not its actual value, which is generically unknown). In the case that GR is the correct theory, $\varphi_k^{\rm T} \rightarrow \varphi_k^{\rm GR}$ and $\delta\hat{\varphi}_k^{\rm T} \rightarrow 0$. Then the systematic error $\Delta(\delta\hat{\varphi}_k)$ measures the shift from 0 caused by binary eccentricity, and $\sigma_{\delta\hat{\varphi}_k}$ is a measure of the fractional measurement precision relative to the true (GR) value of the PN coefficient.  (All of this applies equally well to the log coefficients; we drop that label for simplicity.)
Things are slightly different for the dipole term (Fig.~\ref{errors_dipole}), which vanishes in GR ($\varphi_{-2}^{\rm GR}=0$). In that case $\varphi_{-2}^{\rm T} = \delta\hat{\varphi}_{-2}^{\rm T}$, and the systematic and statistical errors are absolute (not fractional) errors:
\begin{align}
    \Delta(\delta\hat{\varphi}_{-2}) &= \Delta \varphi_{-2} = \varphi_{-2}^{\rm bf} - \varphi_{-2}^{\rm T} \,,\\
    \sigma_{\delta\hat{\varphi}_{-2}} &= \sigma_{\varphi_{-2}} \,.
\end{align}

Considering the statistical errors in Fig. \ref{errors_LIGO}, we note that for BBHs the 0PN, 1PN, 1.5PN, 3PN log, and 3.5PN TGR parameters have a modest measurement precision ($\lesssim 20\%$\mbox{--}$70\%$ error) with a LIGO detector. The 2PN, 2.5PN log, and 3PN coefficients are poorly constrained. For the BNS the 0PN, 1PN, 1.5PN, and 2.5PN log coefficients are measured with a precision of $\approx 1\%$\mbox{--}$7\%$. The 2PN, 3PN, and 3PN log terms are measured with a precision of $\approx 30\%$\mbox{--}$50\%$. This is consistent with the results reported in \cite{GWTC3:2021sio}. For BBHs in CE (Fig.~\ref{errors_CE}), the constraints on the 0PN, 1PN, 1.5PN, 2PN, and 2.5PN log coefficients improve by a factor $\sim 10$ relative to the constraints with LIGO, with the first three coefficients reaching fractional constraints $\sigma_a \sim 0.01$\mbox{--}$0.07$. The 3PN and higher coefficients show a less significant improvement in their precision (by a factor $\lesssim 2$). The BNS statistical errors in the CE band show an improvement by a factor of $\sim 1.3$\mbox{--}$18$ compared to LIGO constraints. 

Constraints on the measurement precision of the $\delta \hat{\varphi}_{-2}$ coefficient (corresponding to dipole radiation) are much stronger. Dipole emission is forbidden in GR but allowed in many alternative theories of gravity \cite{Will:1994fb,Will:2004xi}. Figure \ref{errors_dipole} shows an absolute constraint of $\sim 10^{-6}$\mbox{--}$0.008$ for LIGO. This improves by a factor of order $\sim 100$ for the CE detector. The improved measurement precision of the dipole coefficient arises because it acts as a $-1$PN order term in the GW phase [e.g., a relative $1/v^2 \sim (Mf)^{-2/3}$ correction compared to the leading $0$PN effect]. Statistical errors for the TGR parameters scale like $\sigma_a \sim {1}/{(\rho \partial_a \Psi_{\rm AP})} \sim {1}/{(\rho {\mathcal N}_{k}/\varphi_k)}$, where ${\mathcal N}_{k}$ refers to the number of GW cycles contributed by the $k/2$ PN contribution to the SPA phase \cite{Favata:2021vhw}. In addition to the SNR dependence, the statistical errors in the TGR parameters thus depend on the number of GW cycles \emph{per the magnitude of the PN coefficient} $\varphi_k$. (See Appendix \ref{app:scaling} for a more complete explanation.) For the 0PN term, ${\mathcal N}_0/\varphi_0 \sim {1}/{v_{\rm low}^5} \sim {1}/{(Mf_{\rm low})^{5/3}}$, while ${\mathcal N}_{-2}/\varphi_{-2} \sim {1}/{v_{\rm low}^7} \sim {1}/{(Mf_{\rm low})^{7/3}}$ for the dipole term, where $f_{\rm low}$ is the detector's low-frequency limit. Hence, even though $\varphi_{-2}=0$ in GR, the dipole term potentially contributes more GW cycles (per value of $\varphi_{-2})$ in the detector band, leading to improved measurement precision. Because CE has a smaller value of $f_{\rm low}$, as well as a factor $\sim 10$ improvement in sensitivity relative to LIGO, the measurement precision of $\delta \hat{\varphi}_{-2}$ improves by a factor $\sim 100$.

We next consider the systematic errors. In Fig.~\ref{errors_LIGO} (LIGO) the systematic errors exceed the statistical errors for BBHs at $e_{0} \gtrsim 0.04$, with lower mass BBH systems exceeding the statistical errors at smaller eccentricities. For $M=15 M_{\odot}$, systematic bias is comparable to the statistical error at $e_{0} \approx 0.04$. For heavier BBH systems ($\geq 65 M_{\odot}$), systematic errors do not exceed statistical errors unless $e_0 \gtrsim 0.1$. For the BNS system, systematic errors become greater than statistical errors at smaller values of eccentricity, $\sim 0.008$.

Systematic errors scale according to \cite{Favata:2021vhw}
\begin{equation}
\label{eq:Deltadeltaphi}
    \Delta(\delta \hat{\varphi}_k) \sim e_0^2 \left(\frac{f_0}{f_c}\right)^{19/3} \frac{1}{(Mf_c)^{k/3}} \,.
\end{equation}
Here $f_c$ is a characteristic frequency scale that serves as a dimensional proxy for the actual numerical evaluation of the frequency integrals; see Appendix \ref{app:scaling} for details. Since the SNR scales like $\rho \sim M^{5/6}$, one can then show that the ratio of systematic to statistical errors scales like
\begin{equation}
\label{erroratio}
    \frac{\Delta (\delta \hat{\varphi}_k)}{\sigma_{\delta \hat{\varphi}_k}} \sim \rho {\mathcal N}^{\rm ecc.} \sim \frac{e_0^2}{M^{5/6}} \;,
\end{equation}
where ${\mathcal N}^{\rm ecc.}$ is the number of GW cycles contributed by the eccentric piece of the SPA phase. Hence, heavier mass systems require larger values of $e_0$ for systematic errors to be comparable to statistical errors [$\Delta (\delta \hat{\varphi}_k) \sim \sigma_{\delta \hat{\varphi}_k}$]. For $k>0$, we also see [Eq.~\eqref{eq:Deltadeltaphi}] that at fixed $e_0$ the systematic error decreases with increasing mass, which is the pattern seen in Fig. \ref{errors_LIGO}.

In Fig. \ref{errors_CE} (CE) we see similar behavior. The systematic bias in lower mass BBH systems exceeds the statistical errors at a lower value of $e_0$ compared to higher mass BBH systems. Systematic errors dominate over statistical errors at eccentricities that are a factor $\sim 10$ smaller compared to the LIGO case ($e_{0} \approx 0.005$\mbox{--}$0.01$). For the BNS system, systematic errors cross the statistical errors at $e_0\approx 0.001$\mbox{--}$0.003$.  While systematic errors are independent of SNR, the excellent low-frequency sensitivity of CE (down to $\sim 5$ Hz) allows for higher values of eccentricity in band (for a given value of $e_0$ at $10$ Hz). This produces larger systematic errors for a binary observed with a given $M$ and $e_0$ (in comparison to the systematic error in LIGO). The magnitude of the systematic errors are also many orders of magnitude larger than the statistical errors at higher values of $e_0$. This is due to both the improved low-frequency limit of the detector and its better sensitivity at all frequencies compared to LIGO. The source has a longer inspiral and executes more GW cycles in the CE band. This both lowers the statistical errors (due to the higher SNR) and increases the number of cycles ${\mathcal N}^{\rm ecc.}$ contributed by the eccentric piece of the GW phase [cf., Eq.~\eqref{erroratio}]. CE is thus susceptible to large systematic biases even at very low eccentricities. 

Because they can be measured relatively precisely (small statistical errors), the deformation parameters at 0PN, 1PN, and 1.5PN orders $(\delta \hat{\varphi}_0, \delta \hat{\varphi}_2, \delta \hat{\varphi}_3)$ are most affected by the systematic bias. This systematic error due to eccentricity may lead us to falsely claim a violation of GR. It is therefore essential to include eccentric corrections to the waveform model when constructing meaningful GR tests with data from sensitive 3G detectors.

Returning to Fig. \ref{errors_dipole}, for BBHs we see that the systematic errors on the dipole term become larger than the statistical errors at $e_0 \gtrsim 0.04$ for LIGO. For CE, systematic biases become comparable to statistical errors at $e_0 \gtrsim 0.004$. For the BNS system, the systematic error on the dipole term becomes significant at low values of eccentricity, $e_0\sim 0.008$ and $e_0\sim 0.001$ in the LIGO and CE bands, respectively. The magnitude of the systematic errors on the dipole coefficient is roughly similar for LIGO and CE. In contrast to the non-negative PN order coefficients ($k\geq 0$) shown in Figs. \ref{errors_LIGO} and \ref{errors_CE}, the systematic bias on the dipole term $\delta \hat{\varphi}_{-2}$ is larger for the higher mass systems rather than the lower mass ones. This may be understood via Eq.~\eqref{eq:Deltadeltaphi}, which shows that the systematic bias is proportional to the ratio of eccentric GW cycles to the GW cycles contributed by the $k/2$ PN order piece of the SPA phase (per the magnitude of that term's PN coefficient $\varphi_k$). We see that the ratio changes character for $k<0$. For example, $\Delta(\delta \hat{\varphi}_2) \sim 1/M^{2/3}$ for the 1PN coefficient, but $\Delta(\delta \hat{\varphi}_{-2}) \sim M^{2/3}$ for the -1PN (dipole) coefficient.\footnote{Note that for $k=0$, the systematic bias for the $0$PN parameter $\delta \hat{\varphi}_0$ scales nearly independently with BBH mass. This effect is seen in the $\delta \hat{\varphi}_0$ panel of Fig. \ref{errors_CE}. In Fig. \ref{errors_LIGO} (LIGO) this effect is not seen, likely due to a combination of the additional mass dependence that enters the upper frequency limit $f_{\rm isco}$ and the different low-frequency cutoffs in LIGO and CE noise response.}

In addition to our ``fiducial'' BBHs with parameters $q=2$, $\chi_1=0.5$, and $\chi_2=0.4$, we also explored a handful of BBH systems with other mass ratios and spins. We found that the $e_0$ value corresponding to the intersection point between statistical and systematic errors varies at the level of a few percent to $\sim 25\%$. For example, for a $M=15 M_{\odot}$ binary, changing the value of $q$ from $2$ to $5$ while keeping the spins the same ($\chi_1=0.5$, $\chi_2=0.4$) changes the value of the intersection point for the $\delta \hat{\varphi}_0$ coefficient from $e_0=0.005$ to $0.004$ in the CE band (a $\sim 20\%$ change). In the LIGO band the $e_0$ intersection point changes from $e_0=0.04$ to $0.03$ (a $\sim 25\%$ change).
Changing $q$ from $2$ to $1.2$ results in a negligible change in the $e_0$ intersection value for both the LIGO and CE bands.
 
If we fix $q=2$ but change the spins from $\chi_1=0.5, \chi_2=0.4$ to a maximally aligned configuration ($\chi_1=\chi_2=1$), the intersection point changes from $e_0=0.04$ to $0.041$ in the LIGO band (a $\sim 2.5\%$ change); in the CE band there is a negligible change in the intersection point. If we consider instead a change to a maximally antialigned spin configuration ($\chi_1=\chi_2=-1$), the intersection point changes from $e_0=0.04$ to $0.035$ in the LIGO band (a $\sim 12.5\%$ change). In the CE band the corresponding change is from $e_0=0.005$ to $0.004$ (a $\sim 20\%$ change).


\section{Conclusion \label{sec:conclusion}}
Dynamical formation of BBHs in dense environments such as globular clusters, nuclear star clusters, and active galactic nuclei disks may lead to the formation of highly eccentric binaries. A fraction of these eccentric binaries, when observed in ground-based detectors, may still possess residual eccentricity. Cosmic Explorer is expected to observe $\sim 8.6 \times 10^{4}$ to $5.4 \times 10^{5}$ BBH mergers in a mere $\sim 1$ year of observation time \cite{Baibhav:2019gxm}. With such a high detection rate, it is plausible to expect many eccentric BBH mergers. If quasicircular templates are used for parameter estimation of these eccentric BBHs, it will introduce systematic biases in the estimated TGR parameters even for eccentricities as small as $\approx 0.04$ for LIGO. In the case of  third-generation detectors such as CE, even smaller eccentricities  ($e_0\approx 0.005)$ can lead to systematic biases in the TGR parameters. For BNS the systematic errors become dominant at lower values of eccentricities $e_0\approx 0.008$ in the LIGO band and $e_0\approx 0.002$ in the CE band. Controlling these systematics is extremely important in any search for physics beyond GR. 

In addition to removing systematic bias, waveforms including the effect of orbital eccentricity can also be employed to develop new tests of GR that exploit the eccentric dynamics of these binaries. Considering the non-negligble number of eccentric binaries that could be discovered by 3G detectors, these tests could probe new physics beyond what is considered in the present suite of tests that assume circular binaries.

\section*{Acknowledgements}
We thank Anuradha Gupta for a critical reading of the manuscript and useful comments. K.G.A.~acknowledges support from the Department of Science and Technology and Science and Engineering Research Board (SERB) of India via the following grants: Swarnajayanti Fellowship Grant No. DST/SJF/PSA-01/2017-18, Core Research Grant No. CRG/2021/004565, and MATRICS grant (Mathematical Research Impact Centric Support) No. MTR/2020/000177.
K.G.A and P.S. also acknowledge support from the Infosys Foundation. 
M.F.~was supported by NSF (National Science Foundation) Grant No. PHY-1653374 and a grant from the Simons Foundation (554674, M.F.). This material is based upon work supported by the LIGO Laboratory, a major facility fully funded by the National Science Foundation.
This paper has been assigned the LIGO Preprint No. P2200073.
\appendix
\section{\label{app:scaling}Scaling laws for statistical and systematic errors}
Here we provide a concise summary of scaling laws for statistical and systematic errors that we apply in the interpretation of our results in Sec.~\ref{section4}. The goal is to understand how these errors scale with mass, eccentricity, frequency, and the PN order of a particular TGR parameter. This follows from a similar analysis in Sec.~IV C of \cite{Favata:2021vhw}. See, in addition, Eqs.~(2.19), (4.11), and (4.17) in that reference. Those results demonstrate the following scalings for the statistical and systematic errors for parameter $\theta^a$, as well as for the SNR:
\begin{align}
    \sigma_a &\sim \frac{1}{\rho \partial_a \Psi_{\rm AP}}\,, \label{sigmascale}\\
    \Delta \theta^a &\sim \frac{\Delta \Psi}{\partial_a \Psi_{\rm AP}}  \,, \;\;\;\text{and} \label{sysscale}\\
    \rho &\sim \frac{\eta^{1/2} M^{5/6}}{d_L} \frac{1}{f_c^{1/6} [f_c S_n(f_c)]^{1/2}} \,. \label{SNRscale}
\end{align}
In the SNR scaling we introduce a characteristic frequency scale $f_c$ that preserves the dimensional scaling of the equation; this serves as a proxy for the numerical integration of the SNR integrand over a frequency range $[f_{\rm low}, f_{\rm high}]$.

Considering the form of the circular SPA waveform phasing $\Psi(f)^{\rm circ., TGR}$ in terms of the PN expansion coefficients [Eq.~\eqref{phase} above], we see that the $k/2$ PN term of the phase scales like
\begin{align}
\Psi_{{\rm AP}, k} &\sim \frac{1}{\eta (Mf)^{5/3}}  \varphi_k (Mf)^{k/3} \,, \\
&\sim \frac{\varphi_k}{\eta} \frac{1}{(Mf)^{(5-k)/3}} \,, \nonumber
\end{align}
for $k=(-2, 0,2,3,4,6,7)$. (For simplicity we ignore the log terms in this scaling analysis; it can be easily modified to include those terms.) The eccentric correction that sources the systematic bias $\Delta \Psi$ scales like [Eqs.~\eqref{decomposed phase} and \eqref{eccentric phase}]:
\begin{equation}
\Delta \Psi^{\rm ecc.} = \frac{3\Delta\Psi^{\rm ecc.}_{\rm 3PN}}{128 \eta v^2} \sim -e_0^2\frac{1}{\eta (Mf)^{5/3}} \left( \frac{f_0}{f} \right)^{19/3} \,.
\end{equation}
The derivative with respect to a single TGR parameter scales like
\begin{equation}
\partial_{\delta \hat{\varphi}_k} \Psi_{\rm AP} \sim \frac{1}{\eta} \frac{1}{(Mf)^{(5-k)/3}}\,,
\end{equation}
where $\Psi_{\rm AP} = 2\pi f t_c + \phi_c + \sum_k \Psi_{{\rm AP},k}$.
The corresponding statistical and systematic errors for the TGR parameter $\delta \hat{\varphi}_k$ then scale like
\begin{align}
    \sigma_{\delta \hat{\varphi}_k} &\sim d_L \eta^{1/2} f_c [f_c S_n(f_c)]^{1/2} (M f_c)^{\frac{5-2k}{6}} \,, \\
    \nonumber\\
    \Delta (\delta \hat{\varphi}_k) &\sim e_0^2 \left(\frac{f_0}{f_c}\right)^{19/3} \frac{1}{(Mf_c)^{k/3}} \,.
\end{align}
Hence for $k>0$, we see that the systematic error decreases with increasing total mass, while for $k<0$ (e.g., the dipole term at $k=-2$), the systematic error increases with increasing total mass.

It is also convenient to explain the above scalings in terms of the ``number of cycles'' contributed by a particular PN effect. If we specifically consider the number of GW cycles of the SPA phasing ${\mathcal N}$ [see Eq.~(7.8) of \cite{Moore:2016qxz} for a precise definition], then (roughly speaking) the total number of cycles scales like ${\mathcal N} \sim \Psi_{\rm AP}$. (This must be evaluated as a difference between the end points of a frequency range, but we can consider the result to scale as the function $\Psi_{\rm AP}$ evaluated at a characteristic frequency $f_c$.) The number of cycles contributed by a particular PN term then scales like ${\mathcal N}_k \sim \Psi_{{\rm AP},k}$. In the case of the TGR coefficients (but not necessarily any source parameter), 
\begin{equation}
    {\mathcal N}_k \sim \Psi_{{\rm AP}, k} \sim \varphi_k \, \partial_{\delta \hat{\varphi}_k} \Psi_{\rm AP} \,.
\end{equation}
Then the statistical and systematic errors scale like
\begin{align}
    \sigma_{\delta \hat{\varphi}_k} &\sim \frac{1}{\rho {\mathcal N}_k/\varphi_k} \,, \\
    \nonumber \\
    \Delta (\delta \hat{\varphi}_k) &\sim \frac{{\mathcal N}^{\rm ecc.}}{{\mathcal N}_k/{\varphi_k}} \,,
\end{align}
where ${\mathcal N}^{\rm ecc.} \sim \Delta \Psi^{\rm ecc.}$ is the number of cycles contributed by the eccentric contribution to the SPA phasing. The fractional systematic bias then scales like
\begin{equation}
\frac{\Delta (\delta \hat{\varphi}_k)}{\varphi_k} \sim \frac{{\mathcal N}^{\rm ecc.}}{{\mathcal N}_k} \,,
\end{equation}
and the ratio of systematic to statistical errors scales like
\begin{align}
    \frac{\Delta (\delta \hat{\varphi}_k)}{\sigma_{\delta \hat{\varphi}_k}} &\sim \rho {\mathcal N}^{\rm ecc.} \,, \\
    &\sim e_0^2 \left(\frac{f_0}{f_c}\right)^{19/3} \frac{1}{d_L \sqrt{\eta} f_c^{1/6} \sqrt{f_c S_n}} \frac{M^{5/6}}{(M f_c)^{5/3}} \,, \nonumber \\
    &\sim \frac{e_0^2}{M^{5/6}} \;. \nonumber
\end{align}
Note that the latter ratio does not depend on $k$.
\bibliographystyle{apsrev}
\bibliography{bibliography}
\end{document}